\documentstyle[12pt,aaspp4]{article}

\received{1999 June 14}
\accepted{}
\journalid{}{}
\articleid{}{}

\slugcomment{Draft for ApJ}

\lefthead{Mohanty et al..}
\righthead{Activity in Very Cool Stars:  Magnetic Dissipation in late-M and L Dwarf Atmospheres}

\begin{document}

\def\er{\hat{e}_{r} }
\def\eth{\hat{e}_{\theta} }
\def\ep{\hat{e}_{\phi} }

\def\pr{{\partial}r }
\def\pth{{\partial}{\theta} }
\def\pp{{\partial}{\phi} }
\def\pt{{{\partial}t} }

\def\vn{\vec{v}_n }
\def\veff{\vec{v}_{eff} }
\def\va{\vec{v}_{\alpha} }
\def\vi{\vec{v}_i\ }
\def\ve{\vec{v}_e }
\def\Vnorm{\bar{V} }
\def\vp{v_{\phi} }

\def\pvp{{\partial}\vp }

\def\B{{\vec{B}} }
\def\Bt{{\vec{B_T}} }
\def\Bint{{\vec{B_0}} }
\def\Br{B_r }
\def\Bth{B_{\theta} }
\def\Bp{B_{\phi} }
\def\Brth{B_{r,\theta} }

\def\pB{{\partial}\B }
\def\pBr{{\partial}\Br }
\def\pBth{{\partial}\Bth }
\def\pBp{{\partial}\Bp }

\def\E{\vec{E} }

\def\j{\vec{j} }

\def\Q{\vec{Q} }

\def\Rm{R_m }

\def\Lv{L_{v}}

\def\Lvr{L_{vr}}

\def\Lvth{L_{v\theta} }

\def\LB{L_B }
\def\Leta{L_{\eta} }

\def\dn{{\rho}_n }
\def\dt{{\rho}_{tot} }
\def\di{{\rho}_i }
\def\de{{\rho}_e }
\def\da{{\rho}_{\alpha} }
\def\db{{\rho}_{\beta} }

\def\nn{n_n }
\def\ni{n_i }
\def\ne{n_e }
\def\na{n_{\alpha} }
\def\nb{n_{\beta} }

\def\tni{t_{ni} }
\def\tne{t_{ne} }
\def\tie{t_{ie} }
\def\tab{t_{{\alpha}{\beta}} }

\def\tj{{\tau}_{J} }
\def\tr{{\tau}_{Ross} }

\def\tin{t_{in} }
\def\ten{t_{en} }
\def\tei{t_{ei} }
\def\tba{t_{{\beta}{\alpha}} }

\def\fni{{\nu}_{ni} }
\def\fne{{\nu}_{ne} }
\def\fie{{\nu}_{ie} }
\def\fab{{\nu}_{{\alpha}{\beta}} }
\def\fpe{{\omega}_{pe} }

\def\fin{{\nu}_{in} }
\def\fen{{\nu}_{en} }
\def\fib{{\nu}_{iB} }
\def\feb{{\nu}_{eB} }
\def\fei{{\nu}_{ei} }
\def\fab{{\nu}_{{\alpha}{\beta}} }
\def\fba{{\nu}_{{\beta}{\alpha}} }

\def\ani{a_{ni} }
\def\ane{a_{ne} }
\def\aie{a_{ie} }

\def\mn{m_n }
\def\mi{m_i }
\def\me{m_e }
\def\mh{m_{H_2} }
\def\ma{m_{\alpha} }
\def\mb{m_{\beta} }

\def\dcad{{\eta}_{AD} }
\def\dcd{{\eta}_{d} }
\def\dco{{\eta}_{Ohm} }

\def\r0{r_0 }
\def\r1{r_1 }

\def\fpi{4{\pi} }
\def\ka{\kappa }
\def\si{\sigma }
\def\om{\omega }
\def\siomab{{<{\sigma}v>}_{{\alpha}{\beta}} }
\def\siomba{{<{\sigma}v>}_{{\beta}{\alpha}} }
\def\siomsh{{<{\sigma}v>}_{{{Na}^{+}}{H_2}} }
\def\siomeh{{<{\sigma}v>}_{{e}{H_2}} }

\def\del{\vec{\nabla} }
\def\delsq{{\nabla}^{2} }

\def\hal{H$\alpha$ }
\def\sodi{{Na}$^{+}$ }
\def\hyd{H$_2$ }

\def\kms{km s$^{-1}$\ }
\def\cms{cm s$^{-1}$\ }
\def\cmc{cm$^{-3}$\ }
\def\cmss{cm$^{2}$ s$^{-1}$\ }
\def\cmcs{cm$^{3}$ s$^{-1}$\ }

\def\msun{M$_\odot$\ }
\def\mj{M$_J$\ }
\def\Lhal{L_{H\alpha} }
\def\Lx{L_X }
\def\Lbol{L_{bol} }

\def\Fhal{F_{H\alpha} }

\def\teff{T$_{e\! f\! f}$~}
\def\vsini{{\it v}~sin{\it i}~}

\title{Activity in Very Cool Stars: \nl Magnetic Dissipation in late-M and L Dwarf Atmospheres}

\author{Subhanjoy Mohanty\altaffilmark{1}, Gibor Basri\altaffilmark{1}, Frank Shu\altaffilmark{1}, France Allard\altaffilmark{2}, Gilles Chabrier\altaffilmark{2}}

\altaffiltext{1}{Astronomy Department, University of California, Berkeley, CA 94720.  subu@astron.berkeley.edu; basri@soleil.berkeley.edu; shu@astron.berkeley.edu}
\altaffiltext{2}{CRAL, \'{E}cole Normale Sup\'{e}rieure, 46 Allee d'Italie, Lyon, 69364, France.  fallard@ens-lyon.fr; chabrier@ens-lyon.fr}

\begin{abstract} 
Recent observations show that chromospheric and coronal activity in late-M and L dwarfs is much lower than in the earlier M types.  This is particularly surprising, given that the late-M and L dwarfs are comparatively very rapid rotators:  in the early M dwarfs, rapid rotation is associated with high activity levels.  One possibility is that that the drop-off in activity in the late-M's and L's is a result of very high electrical resistivities in their dense, cool and predominantly neutral atmospheres.

We calculate the magnetic field diffusivity in the atmospheres of objects with \teff in the range 3000 - 1500 K (mid-M to late-L), using the atmospheric structure models of Allard and Hauschildt.  We find that the combination of very low ionization fraction and high density in these atmospheres results in very large resistivities, and thus efficient field diffusion.  While both ambipolar diffusion and Ohmic decay of currents due to ion-electron collisions occur, the primary diffusion effects are due to current decay through collisions of charged particles with neutrals.  Moreover, the latter resistivity is a strong function of both effective temperature and optical depth, increasing rapidly as either \teff or optical depth decreases.  

This has three implications.  One, any magnetic field present is increasingly decoupled from atmospheric fluid motions as one moves from mid-M to L.  In the late-M and L dwarfs, atmospheric motions cannot lead to equilibrium field configurations very different from potential ones.  That is, the magnitude of magnetic stresses generated by atmospheric motions is very small in these objects.  We quantify this effect by a simple Reynolds number calculation.  Two, even if magnetic stresses are easily produced by fluid motions in the hot interior (where the coupling between field and matter is good), their propagation up through the atmosphere will be increasingly hampered by the growing atmospheric resistivity as one moves from mid-M to late L.  Three, these cool dwarfs are expected to be fully convective, with magnetic fields that are generated by a turbulent dynamo.  However, the poor coupling between field and matter in the atmosphere suggests that an efficient dynamo cannot be maintained near the surface, but only at large depths.  Since the turbulent dynamo is thought to produce small-scale fields, burying it at great depths would mean that by the time one reaches the atmosphere, the field strengths would be very weak, and increasingly so with lower \teff (as the dynamo gets pushed further into the interior).  This would exacerbate the difficulty in producing and transporting magnetic stresses in the atmosphere.

In summary, both the generation and propagation of magnetic stresses are increasingly damped with decreasing \teff, in these cool dwarfs.  As a result, the magnetic free energy available for the support of a corona, chromosphere and activity becomes smaller and smaller with later type.  This can account for the observed drop in activity from mid-M to L, {\it assuming} that activity in these dwarfs is magnetically driven.  To check the latter assumption, we estimate the emergent acoustic fluxes in these objects through a Lighthill-Proudman calculation.  While the acoustic fluxes also decrease with decreasing \teff, they appear inadequate to explain the observed \hal fluxes in mid-M to L dwarfs.  In the absence of acoustic heating, magnetic heating indeed seems the most viable way of generating activity.  

Finally, while our calculations do not address flares in late-M and L dwarfs, we speculate that the latter could be created by buoyant flux tubes that are generated in the interior and rise rapidly through the atmosphere, dissipating their associated currents in the highly resistive upper atmospheric layers.    
\end{abstract}


%\clearpage

\section{Introduction}
Recent observations indicate that the ``standard'' rotation-activity connection, observed in stars between roughly F5 to M8.5, breaks down in cooler dwarfs.  While high rotation velocities lead to high chromospheric and coronal activity in the former spectral types, M9 and later dwarfs exhibit very low activity levels in spite of being rapid rotators.  A possible physical basis for this behavior is the topic of this paper.  To begin, we first discuss the usual rotation-activity connection in \S 1.1.  We then examine the evidence for its breakdown in the later dwarfs, and the suggested reasons for this, in \S 1.2.  We note that, for the purposes of this paper, only M dwarfs M9 and later will be referred to as late M; all others are called early or mid-M. 

\subsection{``Standard'' Rotation-Activity Connection}
A connection between rotation and activity is observed in spectral types F5 to M8.  In these objects, faster rotation corresponds to higher levels of chromospheric and coronal activity (as measured by chromospheric activity indicators such as $\Lhal/\Lbol$, and coronal ones such as $\Lx/\Lbol$).  This occurs till a cutoff rotational velocity is reached; for objects rotating even faster, the activity saturates at some maximum value.  Simultaneously, rotation also evolves as a function of age and spectral type in stars later than F5: rotational velocity decreases with age in these objects, but the spindown timescale is longer for later types.  

For stars in the range F5 to M3, these observations are understood through the following paradigm.  These objects have a radiative core and a convective envelope.  Magnetic fields are expected to be generated through the operation of an $\alpha\Omega$ dynamo at the core-envelope interface.  Magnetic stresses are created through the dragging of field lines by photospheric fluid motions; the release of these stresses in the upper atmosphere provides energetic support for a corona and chromosphere and drives activity.  The efficiency of the $\alpha\Omega$ dynamo (i.e, the rate of field generation) rises strongly with decreasing Rossby number ($R$), where $R = P/{{\tau}_c}$, the ratio of the stellar rotation period ($P$) to the convective overturn timescale (${\tau}_c$).  For a given stellar radius and convective timescale, therefore, larger rotational velocities lead to greater dynamo efficiencies and hence higher levels of activity, as observed.  The phenomenon of saturation is not yet well understood.  One proposal is that saturation occurs when the the rotational velocity is sufficiently high to generate fields that cover the entire stellar surface.  Note that, in this $\alpha\Omega$ paradigm, it is not rotational velocity per se, but the Rossby number that is the fundamental parameter.  Indeed, Soderblom et al (1993) find that $\Lhal/\Lbol$ in Pleiades G and K dwarfs is better correlated with Rossby number than with \vsini alone; similar results have been obtained for different samples by other investigators.  

Concurrently, for stars between F5 and M3, angular momentum is extracted from the convective envelope and lost through a magnetized wind, leading to a gradual spindown of the star.  As a result, rotational velocity, magnetic field generation and activity all decrease with time.  The convective envelope deepens as one moves to later types.  Therefore the fraction of stellar mass within the convective envelope, and thus the fractional moment of inertia of the envelope, increase with later type.  Consequently, it takes longer to spin down stars of later types, as observed. 

Stars later than about M3, on the other hand, are fully convective.  The rotation-activity relation observed from M3 to M8 cannot, therefore, be explained by the $\alpha\Omega$ paradigm (the situation later than M3 is analogous to that in stars earlier than F5, except that the $\alpha\Omega$ dynamo fails because the star is fully convective, and not because it is mostly radiative).  In these stars, magnetic fields may also be generated by the ${\alpha}^2$ dynamo (\cite{Radler90}).  The ${\alpha}^2$ dynamo is strongly dependent on Rossby number, so a rotation-activity connection may be expected.  By the argument given earlier relating spindown timescale to the depth of the convection zone, fully convective dwarfs later than M3 may also take longer to spin down than earlier M's, which is consistent with observations.  

Magnetic fields in the M3 to M8 dwarfs may also be generated by a `turbulent' dynamo.  In this picture, the longer spindown timescales of M dwarfs compared to earlier types is a result of the small-scale nature of the magnetic field produced by turbulent dynamos (unlike the large-scale fields from $\alpha\Omega$ and ${\alpha}^2$ dynamos), which makes angular momentum loss more difficult (\cite{Durney}).  However, the efficiency of the turbulent dynamo is only mildly enhanced by faster rotation (\cite{Durney}).  One might expect, therefore, a sharp weakening of the rotation-activity connection at $\sim$ M3, where full convection sets in.  No such change is observed from M0 to M8, however (\cite{Delfosse98}; \cite{Mohanty02}, hereafter MB; \cite{Mohantyprep}):  activity is seen to saturate beyond a cutoff velocity in all these objects, just as in earlier types, and the saturation level (e.g., $\Lhal/\Lbol \sim$ 10$^{-4}$ for chromospheric activity) is similar to that in earlier types.  This may indicate that the ${\alpha}^2$ dynamo actually dominates in these objects.  Alternatively, it may be that the turbulent dynamo is predominantly responsible for field generation even before the onset of full convection, so that the transition to full convection is not observationally significant.  This is not implausible given that this dynamo operates throughout the convection zone, and that even at M0, a substantial fraction of the star is convective.  However, this explanation is not fully satisfactory, given that activity down to about M6 seems to saturate at very small rotational velocities ($\lesssim$ 3 \kms; \cite{Delfosse98}): it is unclear how a weakly rotation-dependent dynamo could induce saturation at such low velocities (unless it creates very strong fields to begin with).  Recently, MB have observed that the threshold velocity for saturation in \hal surface flux ($\Fhal$) may be somewhat higher in M6 and later dwarfs ($\sim$ 10 \kms), and that at lower velocities, no rotation-activity connection is apparent.  This may be attributed to the presence of a weakly rotation-dependent turbulent dynamo:  at low rotational velocities, the field generation rate, and hence activity, is effectively independent of rotation, but at high enough velocities, rotation finally becomes important, causing enhanced field generation and saturated activity.  Whether an ${\alpha}^2$ dynamo, a turbulent one, or both are operational in these stars needs to be clarified by further observations.   

In any case, whatever the details of the field generation process may be in the fully convective stars down to $\sim$ M8, it is clear that these dwarfs have no difficulty in producing substantial activity.  As mentioned above, \cite{Delfosse98} have shown that both saturated and unsaturated activity levels in dwarfs down to $\sim$ M6 are similar to that in the early M dwarfs, and MB have shown that this is true even for dwarfs down to $\sim$ M8 (with rapid rotation, i.e., \vsini $\gtrsim$ 10 \kms, always leading to saturation).  This is in sharp contrast to the situation in M9 and later dwarfs, which we now briefly discuss.

\subsection{Rotation and Activity in Late M and L Dwarfs}
Gizis et al. (2000) report a turnover in the fraction of cool dwarfs with \hal in emission.  The fraction rises from early to mid-M, so that by M7-M8, $\sim$ 100\% of the dwarfs are in emission.  Thereafter, though, the fraction quickly drops, to about 60\% by L0 and 8\% by L4.  None of their sample L5 and later shows definite signs of emission.  This is all the more remarkable, given that the bolometric flux falls sharply with decreasing \teff.  Therefore a constant \hal flux should be increasingly evident with later types against the fainter background.  The observations clearly imply that chromospheric activity falls off after $\sim$ M8\footnote{\hal emission has been seen in a T dwarf (\cite{Burgasser00}), which is much cooler than L dwarfs.  However, most T dwarfs do not show any emission, and this particular case is regarded as an anomaly; the emission is thought to have external causes, such as Roche lobe overflow from a companion, and not an indicator of intrinsic chromospheric activity}.  

Consistent with the Gizis et al. (2000) result, MB report that activity levels (measured either in terms of $\Fhal$ or $\Lhal/\Lbol$) sharply drop beginning $\sim$ M9.  The majority of the MB L dwarfs exhibit no \hal emission.  Moreover, even the maximum activity detected in the L's is barely equal to the minimum levels in the M5 to M8 dwarfs, and more than an order of magnitude lower than the saturated M5-M8 levels.  This is remarkable, especially since the L dwarfs, by M dwarf standards, are very rapid rotators:  almost all the L's in the MB sample have \vsini $>$ 20 \kms, and the fastest has \vsini $\approx$ 60 \kms.  Evidently, the ``standard'' rotation-activity connection breaks down at about M9.

A variety of reasons have been proposed to explain this behavior.  Basri (2000) proposed that very fast rotation may order fluid motions, thereby damping turbulence and lowering the efficiency of any turbulent dynamo.  This no longer seems likely however, given that the L dwarf with highest \vsini exhibits the highest chromospheric activity among the MB L dwarf sample.  Alternatively, models indicate that the convective velocities diminish as one moves from mid-M to the late M and L dwarfs (see Figs. 9 and 10).  Assuming that turbulence is driven by the convective motions, lower convective velocities may imply less energy in turbulent motions and hence a damping of the turbulent dynamo.  The latter cannot be the whole story, however, for three reasons.  

One, the models suggest that convective velocities decrease continuously as one moves from mid-M to later types.  In spite of this, M6 to M8 dwarfs seem quite capable of generating substantial activity.  It is difficult to see why the diminishing convective velocity would suddenly begin to have an observable effect on the dynamo at M9 and not before (see, however, the discussion of acoustic heating in \S 5.3).  Two, dwarfs at M9 (LP 944-20) and M9.5 (BRI 0021) that exhibit little or no basal chromospheric and coronal activity have been observed to flare (\cite{Rutledge00}; \cite{Reid99}).  This implies that whatever dynamo is operational in them is indeed capable of generating substantial fields, but that the energy in the field cannot usually be extracted to drive activity.  Three, as discussed above, an ${\alpha}^2$ dynamo instead of a turbulent one may be dominant in these fully convective dwarfs.  Since the efficiency of this dynamo increases strongly with decreasing Rossby number, its effects should become stronger with decreasing \teff (which in these objects translates to longer convective timescales) and faster rotation, not weaker.  

A different solution has been proposed by Meyer \& Meyer-Hofmeister (1999), which addresses the efficiency of converting magnetic field energy into activity, rather than the efficiency of the field-producing dynamo itself.  They suggest that the electrical resistivities in the cool, dense and predominantly neutral atmospheres of late-M and L dwarfs are very high.  Consequently, these atmospheres cannot support substantial currents.  As a result, since $\del\times\B = (4\pi/c)\j$, large non-potential field configurations (i.e., $\mid\del\times\B\mid \gg 0$) are not possible in the atmosphere.  Potential fields represent the lowest energy state; without large departures from this state, the magnetic free energy available to support a chromosphere and corona and drive activity is small.  {\it It is this scenario that we explore in the present paper}.  

The above picture assumes that magnetic fields are the underlying cause of chromospheric and coronal activity in these cool dwarfs.  It is possible that other energy sources are responsible for activity instead.  The most important candidate in this regard is acoustic energy (which may, for example, play a part in supporting the lower chromosphere of the Sun).  To examine this question, we calculate acoustic fluxes as well in the \teff range of interest here, and compare them to the observed \hal fluxes.

In \S 2, we describe the atmospheric conditions in mid-M to L dwarfs, and the basic relevant equations.  The latter are given in more detail in the Appendix.  In \S 3, we derive the diffusion coefficients in these atmospheres, for a range of optical depths.  The magnitudes of the coefficients are calculated, and the dominant diffusion process found.  Using the latter, we derive the equilibrium field diffusion equation in \S 4.  The consequences of this equation are discussed in \S 5.  In \S 5.1 we calculate the atmospheric Reynolds numbers.  These indicate how well the field is coupled to atmospheric motions, and the magnitude of non-potential field that can be generated by such motions.  We show in \S 5.2 that the Reynolds numbers imply that the observed falloff in \hal activity with \teff may indeed be a consequence of a decline in field-atmosphere coupling.  In \S 5.3, we discuss acoustic waves as an alternative to magnetic fields for generating activity.  Finally, in \S 5.4 we suggest a possible scenario whereby flaring can occur in otherwise inactive ultra-cool dwarfs.  Our results are summarized in \S 6.  

\section{Atmospheric Conditions and Basic Equations}
We first discuss the relevant MHD equations, using conditions obtained in late-M and L dwarf atmospheres.  The atmosphere is assumed to be globally neutral.  Dust grains may be present, but since grain parameters for various processes, e.g., sticking and electron capture, are not very well known, their inclusion in the dynamics of particle interactions is problematic.  Profile-fitting of the observed atomic resonance lines of alkali metals in the optical spectrum of L dwarfs indicates that grains may have gravitationally settled out of the atmosphere (\cite{BasriMoh00})\footnote{The data is best fit by ``Cleared Dust'' models.  Note that the alkali metals themselves do {\it not} form grains; rather, it is the refractory elements such as Ti and V that are depleted.}, and can thus be ignored.  On the other hand, analysis of the infra-red spectrum of late-M and L dwarfs suggests that dust may be present.  A possible solution is that the dust has largely settled out of the upper atmosphere where the optical spectrum is formed, but still lingers in the deeper parts of the atmosphere where the IR lines arise.  In any case, without a good knowledge of grain parameters in these atmospheres, it is not possible to include grains effects in the dynamical equations in a very meaningful way.  We therefore ignore grains in these equations, in the present paper.  On the other hand, models show that backwarming due to dust opacity significantly changes the temperature and ionization fraction profiles at small optical  depths.  We do discuss grain effects, therefore, in determining the environment (temperature, fractional ionization etc.) in which the resistivities are to be evaluated.    

Two classes of models are considered: ``Cleared Dust'' and ``Dusty''.  In both cases, solar metallicity and log[$g$] = 5.0 are assumed.  We consider the effective temperature range \teff = 3000 - 1500 K, corresponding roughly to M5 - L7.  Under these conditions, singly-charged alkali ions constitute the main ionic component (though the alkalis are predominantly neutral in these atmospheres), and \hyd the main neutral component.  Among the alkalis, \sodi ions comprise the dominant ionic species.  The run of various atmospheric parameters with optical depth for a range of \teff is shown in Fig. 1 for ``Cleared Dust'' models, and in Fig. 2 for ``Dusty'' models.  Note the much higher fractional ionization in the ``Dusty'' models compared to the ``Cleared Dust'' ones at low optical depths.  This is due to the backwarming effects of dust.  Regardless of model, we see that the atmospheric fractional ionizations are extremely low over most of the \teff range considered.  Note that all figures are an optical depth scale in the near infra-red J band (with band-center at ${\lambda}_J$ = 1.2$\mu$).    

In Appendix A, we give the MHD equations that are relevant under the circumstances.  Under the assumptions we make  - equilibrium conditions in the motion of the charged particles (i.e, the forces on ions and electrons have come into balance), singly-charged ions, and local thermal equilibrium  - the pressure and gravity terms may be ignored.  The Lorentz force and the drag forces due to collisions completely dominate any other forces acting on the charged particles (see Appendix A for a more complete treatment of this issue).  In the most general case, we would also need to include the fluid equations for the neutrals.  That is, we should take into account the drag on the neutrals due to the magnetic field (via collisions with the charged particles, which directly feel the presence of the field).  Given the high density and low ionization fraction, however, we assume that this is a small correction, which we may ignore in this simplified treatment.  To summarize, we envision neutrals that are put into motion by turbulence and/or convection, and unaffected by the presence of a magnetic field or collisions, and charged particles that move in response to the magnetic field and collisions with other particles (both neutral and charged).  Our task is to examine the behavior of the magnetic field in the presence of such fluid motions.  The MHD equations are solved in Appendix A, to obtain the following result:
$$ \frac{\pB}{\pt} \,\approx\, \del\times(\veff\times\B) \,-\, \del\times\left[(\dcad\,+\dcd\,+\dco)(\del\times\B)\right] \quad \eqno [1a] $$
where,
$$\veff\,\approx\,\vn\,-\,\frac{\j}{e\ni}\,=\,\vn\,-\,(\vi\,-\,\ve) \eqno [1b]$$
Equation [1a] represents the time rate of change in the magnetic field in the presence of fluid motions.  The first term on the R.H.S. of equation [1a] represents the change in the field due to advection by fluid motions, and the second the change in the field due to diffusion and decay processes.  The advection term acts as a source for magnetic stresses, by twisting the field into non-potential configurations.  The diffusion and decay terms, on the other hand, act as sinks, reducing the magnetic stresses.  The equilibrium field (obtained by setting ${\pB}/{\pt}$ to zero in equation [1a]), is determined by the balance between these competing processes.  We now discuss the source and sink terms in more detail.

As shown in equation [1b], the effective velocity of field advection ($\veff$) consists of the neutral velocity ($\vn$), modified by a current term ($\j/e\ni = \vi-\ve$).  The first accounts for field advection by neutrals moving perpendicular to the field, and the second for advection by charged particles moving perpendicular to the field; the latter is the Hall term.  Now, the local current $\j$ only accounts for the {\it non-potential} part of the field, produced in this case by atmospheric fluid motions.  Any background potential field that may be present in the atmosphere is generated by non-local dynamo currents that are present in the interior and do not contribute to the $\j$ considered here.  As we will see, the non-potential component of the field, computed after ignoring the Hall term term, turns out to be very small.  As a result the Hall term is negligible compared to the neutral advection term, and may be safely and self-consistently ignored from the outset (see \S 3 and \S 5.1).  For now, we will continue to write $\veff$ to remind us of the Hall term correction, but assume that $\veff \sim \vn$. 
 
The sink terms are separated into three processes, represented in equation [1a] by three diffusion coefficients:  ambipolar diffusion ($\dcad$), ``decoupled'' diffusion ($\dcd$) and Ohmic diffusion ($\dco$).  Ambipolar diffusion arises when the charged particles are frozen to the field, but the neutrals can flow past it.  $\dcad$ does not represent an actual decay of the field, but the inefficiency of neutral motions in advecting the field and thus twisting it.  ``Decoupled'' and Ohmic diffusion, on the other hand, do represent destruction of the field.  $\dco$ arises from electron-ion collisions, which lead to current decay and thus decay of the associated field.  $\dcd$ is the analogous process resulting from neutral - charged particle collisions.  The current and the associated magnetic field are destroyed as both ions and electrons are knocked off the field lines by collisions with neutrals.  We call this process ``decoupled'' diffusion, since {\it none} of the particles (neutrals, ions, electrons) is frozen to the field if this process dominates - the field is decoupled from particle motions.  We also stress that any basal field produced by interior currents can attain a potential ($\del\times\B$=0), current-free configuration in the atmosphere.  Such a field will not decay through the action of $\dcd$ and $\dco$ in the atmosphere.  The latter account for the destruction of only the twisted ($\del\times\B\neq$0) component of the field, which is superimposed on the basal field and sustained by local atmospheric currents.   

\section{Diffusion Coefficients}
The diffusion coefficients may be written as follows:
$$ {\rm Ambipolar\, diffusion\, coefficient:}\quad\quad \dcad \,\sim\, {{v_A}^2}\tni \,\propto\, \frac{1}{f\cdot{{\dn}^2}} \qquad\qquad\quad\,\, \eqno [2a] $$     
$$ {\rm Decoupled\, diffusion\, coefficient:} \qquad \quad\dcd \,=\, \frac{{c^2}\fen}{{\fpe}^2}\, \,\propto\, \frac{1}{f} \qquad\qquad\qquad\quad \eqno [2b] $$
$$ {\rm Ohmic\, diffusion\, coefficient:}\,\,\,\, \qquad\quad\dco \,=\, \frac{{c^2}\fei}{{\fpe}^2}\,\,\qquad\qquad\qquad\qquad\qquad \eqno [2c] $$

In the above equations, $v_A = \sqrt{{B^2}/4\pi\dt}$ is the Alfv\'{e}n velocity in the combined medium, $\fpe$ is the electron plasma frequency, $f$ is the fractional ionization and $\dn$ the neutral density.  The proportionalities follow from our definitions and the fact that $\dn \sim \dt$ for small fractional ionization.  
 
The Langevin approximation for calculating neutral-charged particle collision timescales is applicable, as long as the slip speed between the neutrals and charged particles is less than the sound speed of the neutrals (\cite{Ciolek93}).  In cool dwarfs (\teff = 3000 - 1500 K), $v_{sound} \sim$ 10$^5$ \cms in the atmosphere.  Convective velocities are much lower, at 10$^3$ to 10$^4$ \cms.  Therefore, if we assume that the slip speed between neutrals and charged particles is at most of the order of the neutral velocity, then the Langevin approximation is valid for neutrals moving at convection speeds.  Moreover, we will see that the neutral densities in these atmospheres are high enough, and the ionization fractions low enough, so that the charged particles are effectively decoupled from the field and swept along with the neutrals.  As a result, the differential velocity between neutrals and ions or electrons is much less than even the neutral velocities.  In other words, we use the Langevin approximation to calculate the diffusion coefficients, and use these to calculate the relative velocity between neutrals and charged particles a posteriori.  If the relative velocity found is smaller than $v_{sound}$, the Langevin approximation may indeed be used self-consistently from the beginning (see end of this section, and \S 5.1).  This is the case over most of the atmosphere for the range of \teff of interest here.  

Under the Langevin approximation, $\si \propto 1/v$, and the combination $\siomba$ is a constant.  For collisions between \sodi and \hyd, $\siomsh$ is $\sim$ 10$^{-9}$ \cmcs (\cite{Ciolek93}).  One can also show, using the Langevin approximation, that $\siomeh \sim {{(\mh/\me)}^{0.5}}\siomsh$.  Using these values, the approximation that $\dn \sim \dt$ for low fractional ionizations, and the values for total density and ionization fraction given by the atmospheric models, we calculate $\fni$ and $\fne$.    Recall also that $\fni$ and $\fne$ are the collision frequencies for a neutral in a sea of ions and a sea of electrons respectively.  The corresponding collision frequency for an ion in a sea of neutrals is $\fin = (\dn/\di)\times\fni$; the collision frequency for an electron in a sea of neutrals is $\fen = (\dn/\de)\times\fne$.  Finally, the collision frequency between ions and electrons is given by $\fie$ (from \cite{Spitzerbook}, adapted for our atmospheric conditions).  The run of $\fin$, $\fen$ and $\fie$ in the atmosphere, for various \teff and both ``Cleared Dust'' and ``Dusty'' models, is given in Figs. 3 and 4.

The gyrofrequency of a singly-charged particle of mass $m$ in a magnetic field $B$ is given by $eB/2{\pi}mc$.  Assuming a B-field on the order of 1 kG gives the gyrofrequency for \sodi ions $\fib \sim$ 7{$\times$}10$^{4}$ Hz, and the gyrofrequency for electrons $\feb \sim$ 3{$\times$}10$^{9}$ Hz.  These are marked in Figs. 3 and 4.  Comparing these gyrofrequencies to the collision frequencies shown in Figs. 3 and 4, we see that $\fin$ $>>$ $\fib$ and $\fen$ $>>$ $\feb$ over most of the atmosphere, i.e., the frequency of collisions with neutrals for both ions and electrons is much larger than their gyrofrequencies.  The implication is that ions and electrons are effectively decoupled from the magnetic field over all but the uppermost atmospheric layers for the range of \teff considered.  Moreover, we also see from Figs. 3 and 4 that the ion-electron collision frequencies are much less than frequency of ion or electron collisions with neutrals.  Therefore we expect the main diffusivity effects to arise from the decoupled diffusion term, given in [2b].  In other words, we expect the Ohmic diffusion term to be completely negligible compared to the decoupling and ambipolar terms, and the decoupling coefficient $\dcd$ to dominate over the ambipolar coefficient $\dcad$ over most of the atmosphere.  Another way to see this is to consider the ratio of $\dcd$ to $\dcad$.  One can easily show that this equals ($\fin$/$\fib$)$\cdot$($\fen$/$\feb$), i.e., the product of the ratios of neutral collision frequency to gyrofrequency for ions and electrons.  $\dcd$ will then dominate completely when the collision frequencies with neutrals, for both ions and electrons, exceed their gyrofrequencies, and $\dcad$ when the situation is reversed.  Over most of the atmosphere, the former condition holds.  Only in the highest reaches of the atmosphere will $\dcad$ dominate.  We verify this by explicitly computing the coefficients (using $B \sim$ 1 kG to calculate $\dcad$).  The run of $\dcd$, $\dcad$ and $\dco$ is shown in Figs. 5 and 6.

The comparison of collision frequencies with gyrofrequencies also provides insight into where the Hall term becomes important, and where the Langevin approximation is valid.  Equation [1b] shows that the Hall term is simply the differential velocity between ions and electrons, which is proportional to the current.  In the situation under consideration, all local currents are ultimately produced by the motion of the neutrals, which, through collisions, drag on the ions and electrons by different amounts.  Intuitively therefore, the greatest differential velocity between ions and electrons will be produced when one of them is efficiently dragged along by the neutrals and effectively decoupled from the field, and the other is largely decoupled from neutral motions and frozen to the field.  In other words, the Hall term will become significant when, for one of the charged species, the momentum-transfer collision frequency with neutrals exceeds the gyrofrequency, and for the oppositely charged species, the gyrofrequency exceeds the collision frequency.  As long as the collision frequency with neutrals for both ions and electrons far exceeds their gyrofrequencies, however, both charged species will be efficiently dragged along by the neutrals and effectively decoupled from the field.  As a result, the differential velocity between ions and electrons, i.e., the Hall term, will remain small compared to the neutral velocity.  Thus the Hall term can be expected to be negligible as long as $\dcd/\dcad\gg$1, i.e., where $\dcd$ dominates.  

Similarly, the Langevin approximation is valid only when the differential velocity between neutrals and charged particles exceeds the neutral sound speed.  This may no longer be true when the charged particles are frozen to the field while the neutrals can freely flow past the field; the ions and electrons will then ``see'' a large neutral velocity.  Conversely, if the charges are decoupled from the field and efficiently collisionally coupled to neutral motions, the differential motion between neutrals and charges will be small, and the Langevin approximation more valid, especially if the neutral velocity itself is much less than $v_{sound}$.  Thus, we can expect the Langevin approximation to hold where $\dcd/\dcad\gg$1, and more so if ${v_n}\ll{v_{sound}}$ in such regions (with ${v_n}$=10$^3$ - 10$^4$ \cms and $v_{sound}\approx$10$^5$ \cms, $\vn\ll{v_{sound}}$ throughout the atmosphere in our case).  In Figs. 5 and 6, we show the diffusion coefficients calculated {\it assuming} the Langevin approximation; by our physical argument here, this assumption is indeed valid wherever $\dcd/\dcad\gg$1.  By our preceding arguments, the Hall term is also negligible in these regions.  These conclusions are corroborated by our Reynolds number calculations in \S 5.1.
     
\section{Linear Field Diffusion Equation in Equilibrium}
In this paper, we are interested in calculating the magnetic stresses produced by large-scale fluid motions in the atmosphere.  We therefore limit our discussion to the regions where such flows may be expected to exist.  The atmospheric models indicate that convection only occurs at $\tj >$ 10$^{-2}$ for the entire range of \teff considered here (3000 - 1500 K), and at $\tj >$ 10$^{-1}$ for all but the hottest (3000 - 2700 K) of these objects.  However, these numbers are probably accurate to only within a pressure scale-height (convection in the models is treated via mixing-length theory).  Besides, convective overshoot can be expected up to a scale-height above the extent of convection itself.  In short, the upper extent of large-scale fluid motions is likely to be $\sim$ 1 scale-height above the upper limit of convection from the models.  This translates to $\tj \gtrsim$ 10$^{-2}$ for \teff of 3000 - 1500 K.  Also, the atmospheric models extend no deeper than $\tj$ = 10$^2$.  In this paper, therefore, we restrict ourselves to the range 10$^{-2}$ $\leq \tj \leq$ 10$^2$.    

Figs. 5 and 6 show that, over this range of optical depths, $\dcd$ is always the dominant diffusion coefficient.  We therefore disregard the ambipolar and Ohmic coefficients henceforth.  This simplifies equation [1a], yielding:
$$ \frac{\pB}{\pt} \,=\, \del\times(\veff\times\B) \,-\, \del\times\left[\dcd(\del\times\B)\right] \eqno [3] $$

We wish to examine the equilibrium solution of equation [3].  After discarding $\pB/\pt$ in equilibrium, expanding out the decay term, and using $\del\cdot\B\,=\,0$ to get $\del\times\del\times\B \,=\, -\, \delsq\B$, equation [3] finally simplifies to:
$$\dcd\delsq\B \,-\, \del\dcd\times(\del\times\B) \,=\, \del\times(\B\times\veff)\eqno [4] $$
This is a linear P.D.E. in $\B$, since $\dcd$ is independent of $\B$, and so is $\veff$ as long as the Hall term is negligible (so that $\veff\approx\vn$, which we assume to be a constant).  We use this equation below to calculate the magnetic Reynolds numbers in the atmospheres of cool dwarfs.    

We stress that this is solely an equilibrium calculation, which is the only regime that is accessible to a simple Reynolds number calculation of the sort undertaken here.  We do not take into account, for example, time-dependent variations in the magnetic field due to changes in the dynamo and the emergence of new flux at the surface.  Such investigations are beyond the scope of this paper, but are necessary in the future to study the temporal behavior of the field more rigorously.       

\section{Discussion}
\subsection{Reynolds numbers}
A precise solution of the equilibrium magnetic field configuration requires a detailed analysis of equation [4].  In this paper, however, we desire only order of magnitude estimates of the magnetic stresses produced by fluid motions.  For this purpose, a dimensional analysis of equation [4] suffices.  Notice also that the second term on the L.H.S. of the equation accounts for the change in the diffusivity coefficient over the extent of fluid motions.  Now, the length-scale of motions due to convection can be assumed to be of order a pressure scale-height.  For \teff = 3000 - 1500 K, this is $\sim$ 10 km in the atmosphere.  On the other hand, we see from Figs. 5 and 6 that $\dcd$ changes by many orders of magnitude over the optical depth range 10$^{-2}$ $\leq \tj \leq$ 10$^2$, for all the \teff considered.  This optical depth range translates to a spatial extent of only 50$\pm$20 km, or a few scale-heights, for both ``Settled Dust'' and ''Dusty'' models (with the hotter objects corresponding to larger spatial extents).  It is therefore clear that the length-scale over which $\dcd$ changes is much smaller than length-scale of convective motions.  In other words, $\dcd$ changes (monotonically with optical depth) much more rapidly than the fluid velocity, and the change in diffusivity cannot be ignored when an exact solution of equation [4] is desired.  However, including $\dcd$ variations greatly complicates the simple dimensional calculation we wish to undertake.  In the following analysis, therefore, we {\it assume} that $\dcd$ is constant over the length-scale of fluid motions.  We then show that our main results are unaffected by rapid variations in $\dcd$.  

With $\dcd$ constant, only the first term on the L.H.S of equation [4] remains to balance the advection term on the R.H.S.  We assume that the initial, background field $\Bint$, which is produced by an interior dynamo, relaxes into a potential (i.e., curl-free) configuration in the atmosphere in the absence of fluid motions.  We further assume that the length-scale over which $\Bint$ varies is much larger than that of fluid motions, so that $\Bint$ can be approximated by a constant over the extent of velocity variations.  Then equation [4] may be written in the following dimensional form:
$$ \frac{{\dcd}{B_T}}{{\LB}^2} \,\approx\, \frac{{B_0}v}{\Lv} \eqno [5] $$
Here, $B_T$ is the magnitude of the twisted, non-potential component of the field, produced by advection of field lines by the fluid.  $\LB$ is the length-scale over which $B_T$ varies, $v$ (= $\mid\veff\mid$) the magnitude of the fluid velocity, and $\Lv$ the length-scale over which $v$ varies ($\sim$ 1 pressure scale-height).  Now, since $\dcd$ is assumed to be constant, variations in the velocity dictate variations in $B_T$, so it is reasonable to suppose that $\LB \sim \Lv$.  Then we finally get:
$$ \frac{B_T}{B_0} \,\approx\, \frac{v\Lv}{\dcd} \,\equiv\, \Rm \eqno [6] $$
Here $\Rm$ is the magnetic Reynolds number.  We see that it is roughly equal to the ratio of the twisted component of the field, produced by fluid motions, to the initial potential field.  For \teff = 3000 - 1500 K, the AH models predict convective velocities of order 10$^3$ to 10$^4$ \cms in the atmosphere (with the higher velocities corresponding to the hotter objects).  Assuming $B_0 \sim$ to 1 kG, $\Lv \sim$ 1 pressure scale-height $\sim$ 10 km and v = 10$^3$ or 10$^4$ \cms, we calculate the Reynolds numbers for 10$^{-2}$ $\leq \tj \leq$ 10$^2$ and \teff = 3000 - 1500 K.\footnote{For a given \teff and optical depth, the AH models predict a particular convective velocity.  However, it is unwise to regard these velocities as precisely the ones that enter into equation [6].  The different spatial components of the convective velocity would give rise to varying magnetic stresses.  Stresses may also be created by convection-driven turbulence, which would have a spectrum of velocities.  Finally, the velocities in the convective overshoot region are not very well known.  For these reasons, we choose to calculate the Reynolds numbers using a fixed convective velocity of either 10$^3$ or 10$^4$ \cms:  according to the AH models, these values bracket the range of velocities that may be expected in the range of optical depths and \teff considered.}  The diffusivity coefficient used is the $\dcd$ at the optical depth where the calculation is performed; as discussed, this assumes that $\dcd$ is constant over the length-scale $\Lv$.  

Before going on to our Reynolds number results, we note that equation [6], which ignores the Hall term, allows us to check the importance of the Hall term and the validity of the Langevin approximation a posteriori.  Assuming ${B_0}\approx$ 1 kG, we calculate $B_T$=${B_0}\Rm$.  Using $(4\pi/c)\j = \del\times\B\approx {B_T}/\Lv$, we calculate $\mid\j\mid$, and thus the Hall term.  Knowing $\mid\j\mid$, $B_0$ and $B_T$ also allows us to calculate $\mid\vi-\vn\mid$ and $\mid\ve-\vn\mid$ from equations [A12a] and [A12b] (where we use $\mid\B\mid\approx {B_0}$, which follows from the fact that the $\Rm$ we derive are, as we shall see below, very small).  From this we can check the validity of the Langevin approximation.  We find that, in agreement with the physical arguments presented in \S 3, the Hall term can indeed be neglected and the Langevin approximation used in the regions where $\dcd$ dominates.  In particular, this holds in the regions of interest here, i.e., those parts of the atmosphere where convective motions may be expected (10$^{-2}$ $\leq \tj \leq$ 10$^2$).

Our Reynolds numbers are plotted in Figs. 7 and 8.  Three important points are immediately apparent, independent of the particular atmospheric model used.  First, the Reynolds number, or equivalently, the magnitude of field twisting produced by atmospheric fluid motions, is extremely low over most of the \teff and optical depth ranges considered.  Only in the hottest of these stars, and deep inside the atmosphere, do the Reynolds numbers become appreciable.  In other words, in the very low ionization fraction and high density atmospheres of these objects, the resistivities due to neutral - charged particle collisions are very high.  Consequently, any existing magnetic field is essentially decoupled from the atmosphere (i.e., from both neutral and charged particles), and fluid motions cannot twist the background field into highly non-potential configurations.  Second, for any given \teff, the Reynolds number decreases with diminishing optical depth:  the resistivity increases with decreasing optical depth, and so the current density that the atmosphere can support correspondingly declines.  Third, at a given optical depth, the Reynolds number decreases with decreasing \teff, reflecting the increase in resistivity in cooler objects.  In other words, the current density that the atmosphere can support at any given optical depth is reduced with later spectral type.

Of course, the Reynolds numbers depend on the velocity and spatial scale of the fluid motions.  Larger velocities and spatial scales will lead to higher Reynolds numbers, since the advection of field lines by the flow becomes more efficient, and stronger non-potential fields can build up.  However, the radial extent of the entire atmosphere in these cool, high-gravity dwarfs is $\lesssim$ 100 km, and v$_{sound}$ about 1 km/s.  Therefore, even in the unlikely case of fluid motions occurring at sound speed and extending over scales comparable to the entire height of the atmosphere in these dwarfs, Figs. 7 and 8 show that the Reynolds numbers at $\tj\lesssim$ 1 would still be less than one, in objects cooler than $\sim$ 1700 K.  That is, below about 1700 K, the atmosphere at and above the photosphere is completely decoupled from the field, even under the best coupling circumstances one might imagine.  

Our results are in agreement with those of Meyer \& Meyer-Hofmeister (1999).  Using their prescription to find the resistivity, one calculates values that are within a factor of two of ours (with ours being the lower of the two), with corresponding agreement in the Reynolds numbers.  The major difference between our studies, with regard to Reynolds numbers, is that their calculations are performed only at the point in the atmosphere where T $\approx$ \teff (i.e., at what may roughly be called the photosphere), and assuming that neutral-charged particle collisions are the dominant resistivity mechanism.  We extend their work by examining the various resistive mechanisms over a range of atmospheric optical depths and for different atmospheric models, and determining the Reynolds numbers over the entire region where convective motions are viable.  This allows one to put the implications for magnetic activity, suggested first by Meyer \& Meyer-Hofmeister and explored in the next section, on a firmer footing.  

Finally, we do not expect our general results to be affected by the rapid change in $\dcd$, for the following reasons.  First, even in the deepest parts of the atmosphere, where the resistivity is {\it lowest}, the Reynolds numbers are still very small ($\Rm <$ 1 at all \teff for v$_{conv}$ = 10$^3$ \cms, and below $\sim$ 2300 K for v$_{conv}$ = 10$^4$ \cms; see Figs. 7 and 8).  $\dcd$ variations do not affect this result since, as Figs. 5 and 6 show, the resistivity changes very slowly at very large optical depths.  Thus, even if the resistivity throughout the atmosphere were equal to the relatively small value at the bottom of the atmosphere, the conclusion that the magnetic field in these cool objects is largely decoupled from atmospheric fluid motions would still be correct.  In reality, the resistivity over most of the atmosphere is much higher than in the deepest regions, which can only strengthen our conclusion.  

This brings us to the second point, which is that the spatial variation in $\dcd$ does not take the form of rapid fluctuations, but merely a rapid monotonic decline with decreasing optical depth.  If the former situation held, it would indeed be incorrect to draw conclusions about the field twisting without accounting for the detailed spatial behavior of $\dcd$.  In the present case however, the resistivities over the length-scale of fluid motions centered on a given optical depth are always smaller than those at a higher optical depth.  Intuitively, this will lead to a decreasing amount of field twisting with smaller optical depth.  This is merely saying that as the resistivity increases with height, the ability of the atmosphere to support large current densities, and hence strong non-potential fields, correspondingly declines.  Of course, the length-scale over which $\dcd$ changes will affect the precise degree of field twisting, so the magnitude of $B_T$ implied by the Reynolds number calculation is likely to be not very precise, but the trend of declining current density with with lower optical depth will remain.

Similarly, we see from Figs. 5 and 6 that, over the range of optical depths where convective motions may be expected ($\tj >$10$^{-2}$), the resistivity at any given optical depth is always higher in a cooler object than in a hotter one.  It is reasonable to conclude therefore, that the current density that the atmosphere can support at a given $\tj$ (and hence the magnitude of non-potential field possible there) declines with effective temperature.  Once again, while the spatial scale of $\dcd$ variation will inform the exact value of $B_T$, the trend with \teff will remain.  

Lastly, if the length-scale $\Leta$ of $\dcd$ variations is much smaller than $\Lv$, then it is likely that $\Leta$ will determine the length-scale $\LB$ over which $B_T$ varies.  If we assume that $\LB \sim \Leta$, then a perusal of equation [5] shows that the ratio ${B_T}/{B_0}$ obtained would be even smaller than what we derive in equation [6] - the field would be even more decoupled from fluid motions.  Moreover, Figs. 5 and 6 show that $\Leta$ becomes smaller (i.e., $\dcd$ declines faster) with both decreasing optical depth (at a given \teff) and decreasing \teff (at a given $\tj$), at least in the regime where convective motions can be expected (both trends are more pronounced in the ``Cleared Dust'' models than in the ``Dusty'' ones).  This behavior would enhance the decline of field twisting with decreasing optical depth and \teff, bolstering our conclusions.  

In summary, while the detailed spatial behavior of the resistivity surely affects the precise magnitude of field twisting produced by atmospheric fluid motions, a simple Reynolds number calculation suffices to elucidate the general features of the atmosphere-field coupling in these cool objects.  In the next section, we discuss the implications of our Reynolds number calculations for magnetically driven activity.  

\subsection{Implications for Magnetically Driven Activity}
In the standard paradigm for explaining activity, the photospheric footpoints of magnetic field lines are advected around by fluid motions.  This twists up the field lines, creating magnetic stresses that are transported to the upper reaches of the atmosphere (by, for example, MHD waves).  There the stresses build up, until they are dissipated through various mechanisms, such as reconnection events and dissipation of MHD waves.  The energy released in the process supports the chromosphere and corona and drives activity.

In stars later than about M5 (i.e., \teff $\lesssim$ 3000 K), however, our Reynolds number calculations show that the photosphere ($\tj \sim$ 1) is only weakly coupled to the magnetic field.  As a result, the magnitude of magnetic stresses that can be generated through field advection by photospheric fluid motions is very small.  Moreover, the Reynolds number at the photosphere drops drastically with decreasing \teff, so smaller and smaller photospheric stresses will be created with later spectral type.  Furthermore, to drive \hal activity, these stresses must be transported to, and dissipated in, the upper atmosphere; dissipating them at large optical depths will not lead to significant heating (and hence \hal emission), since the heat capacity of the deep atmosphere is very high.  However, we have seen that the resistivity of the atmosphere increases strongly with decreasing optical depth, and that this trend becomes more and more pronounced with decreasing \teff.  Consequently, a large part of the magnetic stresses generated in the photosphere will be dissipated before reaching the upper atmosphere (due to the decay of the associated currents along the way), and this effect will be enhanced with later spectral type.  The net result of all this is a sharp decrease, as one moves from mid-M to late-M and L, in the magnetic free energy that is derived from photospheric motions and available for supporting a chromosphere and inciting \hal activity.  

Strong magnetic stresses can of course be generated in the deep atmospheric layers and the interior, where the field is strongly coupled to the matter (as it must be, for a dynamo to efficiently create any magnetic field in the first place).  However, as in the photospheric case discussed above, these stresses must be carried to and dissipated in the upper reaches of the atmosphere, to produce any observable \hal emission.  As before, the associated currents must then traverse an atmosphere that is increasingly resistive with height, and more so at later spectral types.  Much of the current will therefore decay before reaching the upper layers.  Consequently, the magnetic free energy extracted from interior fluid motions and available to drive \hal activity rapidly decreases from mid-M to L.

Finally, mid-M to L dwarfs are expected to be fully convective, with magnetic fields perhaps generated by turbulent dynamos.  However, our Reynolds numbers calculations show that such a dynamo cannot operate efficiently close to the surface, since the resistivities here are very high.  Any efficient dynamo must be buried deep in the interior, where the matter-field coupling is good.  At the same time, turbulent dynamos are expected to produce small-scale fields (of order the spatial scale of the turbulence).  Consequently, if the dynamo is buried at great depths, the field strength by the time one reaches the atmosphere may be very small.  Weak atmospheric fields would only magnify the difficulties, discussed above, in generating and transporting currents in the atmosphere.  This effect may also be exacerbated with later type:  as \teff declines, the region of high resistivity extends further into the interior, the dynamo gets buried at increasingly large depths, and weaker and weaker atmospheric fields result.  \footnote{Note that, if an ${\alpha}^2$ dynamo dominates instead, this mechanism may become invalid, since such a dynamo can generate large-scale fields}.
 
To summarize, our Reynolds number calculations indicate that both the generation and the propagation of magnetic stresses is increasingly damped, with later type, in the highly resistive atmospheres of these cool dwarfs.  This can lead to the decline in \hal emission observed in going from mid-M to late-M and L dwarfs.  This assumes, of course, that \hal activity in these objects is indeed magnetically driven.  To verify this assumption, one needs to compute the emergent magnetic energy fluxes, and compare them to the observed \hal levels.  This requires a detailed calculation of the magnetic field configuration, field reconnection rates, MHD wave damping etc., and is beyond the scope of this paper.  However, one can indirectly investigate this question by asking what non-magnetic energy source can plausibly drive \hal activity.  Acoustic waves are the most viable alternative, and we discuss them below.

\subsection{Acoustic Fluxes}
Ulmschneider et al (1996) investigate acoustic fluxes in late-type stars.  They use mixing length theory  (MLT) to calculate convection characteristics, and calculate the resulting acoustic fluxes using a detailed Lighthill-Stein formulation.  They find that in the late-M dwarfs, acoustic fluxes appear to be insufficient to explain the observed chromospheric activity levels.  They also find that a simple Lighthill-Proudman calculation generally gives acoustic fluxes in good agreement with those derived from the detailed theory.  The agreement worsens with decreasing \teff, with the Lighthill-Proudman fluxes always being higher than the Lighthill-Stein ones.  However, the calculations of Ulmschneider et al are carried out under the assumption of grey opacities in the atmosphere.  This is a poor approximation of the true situation in late-M and L dwarfs, where the presence of a large number of molecular species (e.g. TiO, VO, H$_2$O) and dust creates decidedly non-grey opacities.  Ulmschneider et al speculate that this may lead to higher convective velocities, and hence larger acoustic fluxes in these objects than what they calculate.  

We re-examine the issue here, by applying the Lighthill-Proudman formulation to the convective velocities given in the AH atmospheric models.  These models include a detailed treatment of the non-grey opacities in very cool dwarfs, and produce emergent spectra in good agreement with observations.  The models also calculate convective velocities in the non-grey atmosphere, using MLT with $\alpha$ = 1 (where $\alpha \equiv l/{H_p}$, with $l$ being the mixing length and $H_p$ the pressure scale-height).  

In the Lighthill-Proudman formulation (Ulmschneider et al, 1996), the total acoustic flux emergent from convection layers is, in cgs units:
$$ F_{ac} \,\approx\, \frac{1}{2}\int\limits_{{\Delta}z} 38\rho\frac{{v_c}^8}{{{c_s}^5}{H_p}}dz \eqno[7] $$
Here $\rho$ is the density, $v_c$ the convective velocity, $c_s$ the sound speed and $H_p$ the pressure scale height, at a given point $z$ in the convection zone.  Since $F_{ac}$ goes as the eighth power of $v_c$, it is necessary to perform the integration only over the region $\Delta z$ where the convective velocity is appreciable.  The factor of 1/2 in the above equation accounts for the fact that only half of the total acoustic flux generated is directed outwards.  

In the left-hand panels of Figs. 9 and 10, we show the AH model convective velocities, for ``Settled Dust'' and ``Dusty'' models respectively.  In the right-hand panels of Figs. 9 and 10, we plot the resulting emergent acoustic fluxes using the Lighthill-Proudman formula.  Since $F_{ac}$ goes as ${v_c}^8$, most of the contribution to $F_{ac}$ comes from the peak convective velocity.  Our results are very similar to those of Ulmschneider et al (1996), even though our calculations are done under non-grey conditions.  These acoustic fluxes appear unable to explain both the saturated \hal levels in M dwarfs earlier than M9, and the \hal levels in those L dwarfs where \hal is detected.  In the M8.5 and earlier dwarfs, MB report that \hal emission saturates at $\Fhal \sim$ 10$^{5.5}$, while we find $F_{ac} \sim$ 10$^{2.5}$, or a factor of 10$^3$ lower.  In those L dwarfs where \hal is detected, MB find $\Fhal \sim$ 10$^{3.5}$, while our acoustic fluxes are lower by a factor of 10$^4$ to 10$^5$.  To match the observed $\Fhal$, we need $v_c$ higher by a factor of $\sim$ 2.5 in the M dwarfs (to increase $F_{ac}$ by a factor of 10$^3$), and $v_c$ higher by a factor of $\sim$ 4 in the L dwarfs (to increase $F_{ac}$ by a factor of 10$^5$).  Now, in standard MLT, ${v_c} \propto \alpha$.  Thus, for acoustic fluxes to equal the observed \hal fluxes, we need $\alpha \sim$ 2.5 in M dwarfs and $\alpha \sim$ 4 in the L dwarfs.  Conventionally (and for reproducing solar observations), an $\alpha$ between 1 and 2 is preferred (as is intuitively plausible, since the mixing length parameter $\alpha$ specifies how many pressure scale-heights a rising and expanding blob of fluid traverses before finally dissolving into its surroundings).  An $\alpha$ between 2.5 and 4 is thus a significant departure from standard MLT.  It is true that the MLT is likely to be inaccurate near the photosphere, where radiative effects become felt and convection begins to become inefficient.  However, whether this can lead to an effective $\alpha$ parameter so different from the conventional one is not clear.   

We conclude, therefore, that acoustic waves alone appear unlikely to energetically support a chromosphere, at least in those M dwarfs where the \hal flux is saturated, and in the L dwarfs where \hal emission is observed.  It is possible that acoustic heating does play a significant role in the unsaturated M dwarfs, and in the L dwarfs where \hal is not detected at our observation limits (and for which, therefore, we have no indication of the true levels of chromospheric activity, if any).  As MB report, however, there is a decline in the {\it maximum} observed $\Fhal$ from mid-M to L; our foregoing arguments imply that this is probably not due solely to the decline in acoustic fluxes.  The implication is that any \hal emission observed in these objects is probably largely magnetically driven.  Therefore, the decline in \hal activity seen from mid-M to L may indeed be a consequence of the increasing atmospheric resistivity over this spectral range.  

At the same time, note that, just like the observed \hal flux, the predicted acoustic flux also declines with \teff.  This is a direct consequence of the decline in model convective velocities with \teff (Figs. 9 and 10).  Therefore, even if the acoustic flux contributes to the observed \hal emission, it will serve to magnify the decline in emission  over what would result from a decrease in magnetic heating alone, as one goes to lower \teff. In any case, the usual argument against acoustic fluxes applies here: all stars of the same \teff should look similar if acoustic heating is dominant, and they don't.

\subsection{Flares} 
Finally, we note that the large atmospheric resistivities also make it difficult to create flares in the classical fashion, wherein large magnetic stresses are built up in the upper atmosphere through motions of the field  footpoints, and then released through very rapid reconnection events.  We have seen that such stresses are hard to build up in the upper atmosphere.  However, flares are observed in late M dwarfs, even in those that do not show signs of long-term activity (LP 944-20 being one recent example; \cite{Rutledge00}).  We speculate that such flares could conceivably be generated by strongly twisted flux tubes, produced in the interior (where the conductivity is large enough to produce large magnetic stresses), that rise up rapidly enough through the atmosphere to release their magnetic energy in the highly resistive upper atmospheric layers (we thank Tom Abel for initially suggesting this mechanism to us in private communications, 2001).  It is necessary to transport the stresses to the uppermost layers before completely dissipating them because the heat capacity of the atmosphere in the deeper layers is too high for any heating (and thus, flaring) to occur even if a substantial amount of magnetic energy is dumped there. 

For magnetic stresses to be transported from the interior to the upper layers by buoyant flux tubes, two dissipation mechanisms need to be countered as the tube rises through the bulk of the cool, resistive atmosphere.  The first is recombination - the gas within the tube needs to remain ionized while surrounded by much cooler material outside the tube, otherwise the currents sustaining the stresses will be lost before the upper layers are reached.  The second is collisions with neutrals - the atmosphere is very resistive, as we have shown, and collisions with neutrals can dissipate the currents, again before the tube gets to the highest layers.  To counter these difficulties, a sufficiently thick flux tube is needed.  The gas deep within such a tube can be sufficiently  optically thick to only slowly lose the temperature and ionization state it had in the deep interior.  Neutrals will also take a long time to eat their way, through collisions, to the interior of a thick enough tube. In other words, for this mechanism to work, the rise time for the flux tube has to be shorter than the timescales for current dissipation (through cooling and neutral collisions). If the tube is thick because it has stronger fields, it might also be more buoyant. A thick flux tube is therefore advantageous because it lengthens the dissipation times, and might shorten the rise time.   

It is plausible that sufficiently thick tubes might only be created and rise sporadically. Thin tubes, though generated more easily, would  rapidly decay before rising very far up through the resistive atmosphere, and not lead to any activity or flaring.  This could be one way of generating periodic flares without continuous, long-term activity.  Whether this mechanism can actually work will have to be the subject of further research; at the moment it is only the germ of an idea. Other ideas should be generated and explored as well.

The recent observations by \cite{Berger01} of relatively common radio flaring in ultracool stars make this work more urgent. The fact that the radio flares are not accompanied by the expected level of X-rays probably means that the flare mechanism is not a close analog of solar flares. His suggestion that these flares imply that decoupling of magnetic fields is not occurring in ultracool photospheres is, however, an unwarranted extension of enigmatic observations of phenomena in the outer atmosphere down to the much better understood photosphere. There is no question that the photospheres of ultracool dwarfs are quite neutral, and no question that just above their photospheres there is not much chromospheric (6000-10000K) gas. We can all agree, in any case, that the ultracool dwarfs have opened several fascinating new lines of inquiry in the subject of stellar magnetic activity.

\section{Conclusions}
Observations show that \hal activity rapidly declines from mid-M to L dwarfs.  We have investigated the possibility that this results from high electrical resistivities in the cool, mostly neutral atmospheres of these objects.  

Using the atmospheric models of Allard and Hauschildt, the resistivities are calculated for a range of optical depths, and effective temperatures appropriate to  $\sim$ M5 to L6 dwarfs.  We find that the resistivities are indeed very large, predominantly due to the large rate of collisions between neutrals and charged particles.  The collisions efficiently knock the charged particles off the magnetic field lines, and effectively decouple the field from atmospheric fluid motions.  As a result, fluid motions in the atmosphere cannot generate substantial magnetic stresses.  Since large currents be sustained in this very resistive environment, magnetic stresses generated in the highly conducting interior cannot efficiently traverse the atmosphere either.  Finally, the atmospheric magnetic fields themselves may be weak if a turbulent dynamo is present: the large atmospheric resistivities would push the dynamo into the interior, and the small-scale fields it generates would not reach very far into the atmosphere.  This would further hamper the creation and transport of currents through the atmosphere.  Consequently, both the production and propagation of magnetic stresses is severely hampered in these dwarfs.  Since the atmospheric resistivities increase strongly with decreasing \teff, these difficulties in generation and transport of currents increase as well as \teff goes down.  Therefore the magnetic energy available to support a chromosphere and generate \hal activity strongly diminishes as one moves from mid-M to L.  Hence, the observed decline in \hal activity from mid-M to L may indeed be a consequence of the high atmospheric resistivities.

Our above conclusion assumes that \hal activity is magnetically driven at spectral types M5 and later.  We cannot be certain of the validity of this assumption, since we have not calculated the actual magnetic energy flux resulting from atmospheric and interior fluid motions, and compared it to the observed \hal fluxes.  However, in the absence of magnetic heating, acoustic waves appear to be the only viable energy source for \hal emission.  A simple Lighthill-Proudman calculation shows, that acoustic heating is probably not sufficient to explain the peak \hal fluxes observed in mid-M to L dwarfs, although it may still be energetically important in the unsaturated M's, and those L's in which no \hal emission is currently detected.  Since there is a decline in the peak \hal emission from mid-M to L dwarfs, magnetic heating probably does need to be invoked.  Even if the acoustic flux contributes to the \hal emission, it too decreases with \teff.  Hence it may be expected to further exaggerate the decline in emission, as one moves to cooler dwarfs, that would result from a lessening of magnetic heating alone.  

We have also sketched a possible mechanism whereby flaring might occur even in the absence of continuous, long-term activity.  Our scenario involves thick, twisted flux tubes that are generated in the interior and rapidly rise up and dissipate their energy in the upper atmosphere.  It remains to be seen whether this is indeed a viable mechanism.

\clearpage

\appendix
\section{MHD Equations}
 
$$ e{\ni}(\E + {\frac{\vi}{c}}{\times}\B) \,+\, {\frac{\di}{\tin}}({\vn}-{\vi}) \,+\, {\frac{\di}{\tie}}({\ve}-{\vi}) \,=\, 0 \eqno [A1] $$
$$ -e{\ne}(\E + {\frac{\ve}{c}}{\times}\B) \,+\, {\frac{\de}{\ten}}({\vn}-{\ve}) \,+\, {\frac{\de}{\tei}}({\vi}-{\ve}) \,=\, 0 \,\,\,\eqno [A2] $$
$$ \j \,=\, e({\ni}{\vi}-{\ne}{\ve}) \eqno [A3]$$
$$ \ne \,=\, \ni \eqno [A4]$$
$$ \j \,=\, {\frac{c}{\fpi}}(\del{\times}\B) \eqno [A5]$$
$$ {\del}{\times}{\E} \,=\, -{\frac{1}{c}}{\frac{\pB}{\pt}} \eqno [A6]$$
$$ {\del}{\bf{\cdot}}{\B} \,=\, 0 \eqno [A7] $$
$$ \tab \,=\, {\frac{{\ma}+{\mb}}{\mb}}{\frac{1}{{\nb}{\siomba}}} \eqno [A8] $$
$$ {\frac{\da}{\tab}} \,=\, {\frac{\db}{\tba}} \eqno [A9] $$

Here, the subscript $n$ refers to neutrals, $i$ to ions and $e$ to electrons.  $\ma$, $\na$, $\da$ and $\va$ represent the particle mass, number density, mass density and macroscopic velocity of any species $\alpha$ (neutrals, ions or electrons).  

Equation [A1] and [A2] are the equations of motion for ions and electrons respectively, assuming force balance, i.e, in equilibrium.  We have ignored the pressure and gravity terms in [A1] and [A2]; the justification for this is given at the end of this Appendix.  Given singly-charged ions, eqns. [A3] and [A4] give the current density and express global charge neutrality respectively.  Equation [A5] is Ampere's law after disregarding time-variations in $\E$, and equation [A6] is Faraday's law of induction.  The last two equations in the set are valid for elastic collisions.  Equation [A8] defines $\tab$, the mean (momentum exchange) collision time for a species $\alpha$ in a sea of species $\beta$.  The corresponding collision frequency is defined by $\fab \equiv 1/{\tab}$.  Here $\siomba$ is the average collisional rate between particles of species $\alpha$ and $\beta$.  Equation [A9] is implied by Newton's 3$^{rd}$ law.  In equations [A1] and [A2], the pressure and gravity terms have been left out.  The justification for this is given at the end of this Appendix.

To simplify the equations, we define the following convenient shorthand variables (without attaching any particular physical significance to them):
$$ {\ka}_n \,\equiv\, \frac{\dn}{e\ni} ; \quad {\ka}_i \,\equiv\, \frac{\di}{e\ni} \eqno [A10a] $$
$$ \ani \,\equiv\, {\ka}_n\fni ; \quad \ane \,\equiv\, {\ka}_n\fne ; \quad \aie \,\equiv\, {\ka}_i\fie \eqno [A10b]$$
Then, adding eqns. [A1] and [A2], and using the other relationships, one gets:
$$ \frac{1}{c}\left(\j\times\B\right) \,+\, {\frac{\di}{\tin}}({\vn}-{\vi}) \,+\, {\frac{\de}{\ten}}({\vn}-{\ve}) \,=\, 0 \eqno [A11] $$
which is the force equation for the combined medium of ions and electrons.  The collision terms between ions and electrons have dropped out, numerically because ${\di}/{\tie} = {\de}/{\tei}$ (see [A9]) , and physically because such collisions leave the total momentum of the combined medium unchanged.  We can manipulate this equation to give the relative velocity between the neutrals and either ions or electrons.  These are (using the shorthand notation given in [A10a-b]): 
$$ (\vi \,-\, \vn) \,=\, \frac{1}{e\ni(\ani + \ane)} \cdot \left[ \frac{1}{c}\left(\j\times\B\right) \,+\, \ane\j \right] \eqno [A12a] $$
or equivalently,
$$ (\ve \,-\, \vn) \,=\, \frac{1}{e\ni(\ani + \ane)} \cdot \left[ \frac{1}{c}\left(\j\times\B\right) \,-\, \ani\j \right] \eqno [A12b] $$
Subtracting equation [A2] from equation [A1], dividing throughout by $e\ni$, using the other relationships, and taking the curl of the result, yields:
$$ \frac{\pB}{\pt} \,+\, \del\times(\B\times\vi) \,=\, -\del\times\left[\frac{\ani}{e\ni(\ani + \ane)}\cdot\left(\j\times\B\right)\right] $$
$$ \qquad\qquad\qquad\qquad\qquad\qquad\qquad\qquad\qquad\,\, -\del\times\left[\frac{c\ani\ane}{e\ni(\ani + \ane)}\cdot\j\right] \,-\, \del\times\left[\frac{c\aie}{e\ni}\cdot\j\right] \eqno [A13] $$
If the R.H.S. of the equation [A13] were zero, we would get the usual field-freezing equation.  Under the circumstances, the right-hand terms represent field diffusion due to collisions between the various species (neutrals, ions and electrons).  Using equation [A12a] to eliminate $\vi$ from equation [A13], we get the drift of the magnetic field relative to the neutrals:
$$ \frac{\pB}{\pt} \,+\, \del\times(\B\times\vn) \,=\, -\del\times\left[\frac{\ani-\ane}{e\ni(\ani + \ane)}\cdot\left(\j\times\B\right)\right] $$
$$ \qquad\qquad\qquad\qquad\qquad\qquad\qquad\,\,\, -\del\times\left[\frac{1}{ce\ni(\ani + \ane)}\cdot\left(\B\times\left(\j\times\B\right)\right)\right] $$
$$ \qquad\qquad\qquad\qquad\qquad\qquad\qquad\qquad\qquad\,\, -\del\times\left[\frac{c\ani\ane}{e\ni(\ani + \ane)}\cdot\j\right] \,-\, \del\times\left[\frac{c\aie}{e\ni}\cdot\j\right] \eqno [A14] $$

Transposing the first term on the R.H.S. of equation [A14] to the L.H.S., we finally get:
\clearpage
$$ \frac{\pB}{\pt} \,+\, \del\times(\B\times\veff) \,=\, -\del\times\left[\frac{1}{ce\ni(\ani + \ane)}\cdot\left(\B\times\left(\j\times\B\right)\right)\right] $$
$$ \qquad\qquad\qquad\qquad\qquad\qquad\qquad\qquad\quad -\del\times\left[\frac{c\ani\ane}{e\ni(\ani + \ane)}\cdot\j\right] \,-\, \del\times\left[\frac{c\aie}{e\ni}\cdot\j\right] \eqno [A15a] $$
where,
$$\veff\,\equiv\,\vn\,-\,\frac{\ani - \ane}{e\ni(\ani + \ane)}\j  \eqno [A15b]$$
The first term on the R.H.S accounts for ambipolar diffusion.  The second term accounts for field decay through neutral - charged particle collisions, and the third for field decay through ion - electron collisions.  Using $\j = (c/4\pi)\del\times\B$, and defining the diffusion coefficients:
$$\dcd \,\equiv\, \frac{{c^2}\ani\ane}{4\pi e\ni(\ani+\ane)} \,;\, \dco \,\equiv\, \frac{{c^2}\aie}{4\pi e\ni} \eqno [A16] $$
we get:
$$ \frac{\pB}{\pt} \,+\, \del\times(\B\times\veff) \,=\, -\del\times\left[\frac{\left(\B\times\left(\j\times\B\right)\right)}{ce\ni(\ani + \ane)}\right] \,-\, \del\times\left[(\dcd + \dco)(\del\times\B)\right] \eqno [A17] $$
The first term on the R.H.S., arising from ambipolar diffusion, cannot be exactly reduced to the form of the second term, resulting from field decay due to collisions.  One may do so approximately, however, by using dimensional analysis to define an approximate ambipolar diffusion coefficient:
$$\dcad \,\approx\, \frac{B^2}{4\pi e\ni(\ani+\ane)} \eqno [A18] $$
which implies:
$$ \frac{\pB}{\pt} \,+\, \del\times(\B\times\veff) \,\approx\, - \del\times\left[(\dcad + \dcd + \dco)(\del\times\B)\right] \eqno [A19] $$ 
This is the field diffusion equation given in equation [1a] of \S 2.  This is a non-linear P.D.E. in $\B$, due to the $B$-dependence of $\veff$ and $\dcad$.  Substituting back the expressions (equations [A10] and [A11]) for ${\ka}_n$, ${\ka}_i$, $\ani$, $\ane$ and $\aie$, and noting that in our case the mass of the ions (\sodi) greatly exceeds that of the neutrals (\hyd) and the electrons, transforms equation [A15b] into [1b] of \S 2, and equations [A18] and [A16] into equations [2a-c] of \S 3.

Finally, we return to the issue of the pressure and gravity forces.  To include them in the equations of motion of the ions and electrons, we need to add the terms $-\del{p_i}$ and $-\di\del\phi$ to the L.H.S. of [A1], and $-\del{p_e}$ and $-\de\del\phi$ to the L.H.S. of [A2].  $p_i$ and $p_e$ represent the local isotropic pressure due to ions and electrons respectively, and $\phi$ the gravitational potential.  Now, with singly-charged ions and a globally neutral medium, we have $\ni = \ne$.  Combined with the assumption of local thermal equilibrium, this implies that ${p_i} = {p_e}$, and so $\del{p_i} = \del{p_e}$.  Then, on adding [A1] and [A2], we get the additional terms $-\left[2\del{p_i} + (\di + \de)\del\phi\right]$ on the L.H.S. of [A11] (i.e., in the force equation for the combined medium of charged particles).  On subtracting [A2] from [A1], the pressure terms cancel out (a result of the fact that we assume force balance for the ions and electrons individually, as well as singly-charged ions, global neutrality and local thermal equilibrium).  The gravity term remains, after dividing throughout by $e\ni$, as $\left[({m_i} + {m_e})/e\right]\del\phi$, but drops out as well when we take the curl to get [A13], the evolutionary equation for the magnetic field (since $m_i$,$m_e$ and $e$ are constants, and $\del\times\del\phi = 0$).  In other words, under the physical conditions we assume, neither the pressure nor the gravity terms matter insofar as the evolution of the magnetic field is concerned.  They do enter into the force equation for the combined charged medium, however.

To solve the force equation, [A11], with the gravity and pressure terms included, is highly complicated.  Instead, we take the following route.  We solve our system of equations as detailed previously, by ignoring the pressure and gravity terms in [A11], and calculate the resulting equilibrium magnitudes of the Lorentz force and drag forces due to collisions with neutrals\footnote{Equation [6] in the main text is used to calculate $B_T/\Lv$, assuming a basal field $B_0$ = 1 kG and $v$ = 10$^3$ or 10$^4$ \cms, as discussed in the main text.  $j$ is calculated to order of magnitude as $(c/4\pi)({B_T}/\Lv)$, and [A12a-b] are used to calculate the magnitude of the differential velocity between neutrals and charged particles.  From these quantities, the magnitude of the Lorentz and drag forces are estimated.}.  These are then compared to the pressure and gravity terms (assuming $g \approx$ 10$^5$ cms$^{-2}$).  In all cases, the latter two forces are found to completely negligible, by many orders of magnitude, compared to the Lorentz and drag forces.  Therefore we may self-consistently ignore the pressure and gravity terms in the force equation from the outset.  Since they do not enter in the field-evolution equation either, they may be dropped altogether from equations [A1] and [A2].   

\acknowledgements
\section{Acknowledgements}
SM would like to thank Tom Abel for insightful discussions and suggestions on flares and rising flux tubes. Support for this research was derived from NSF grant AST-96-18439 and Chandra grant SAO GO0-1009X.

\clearpage

\plotone{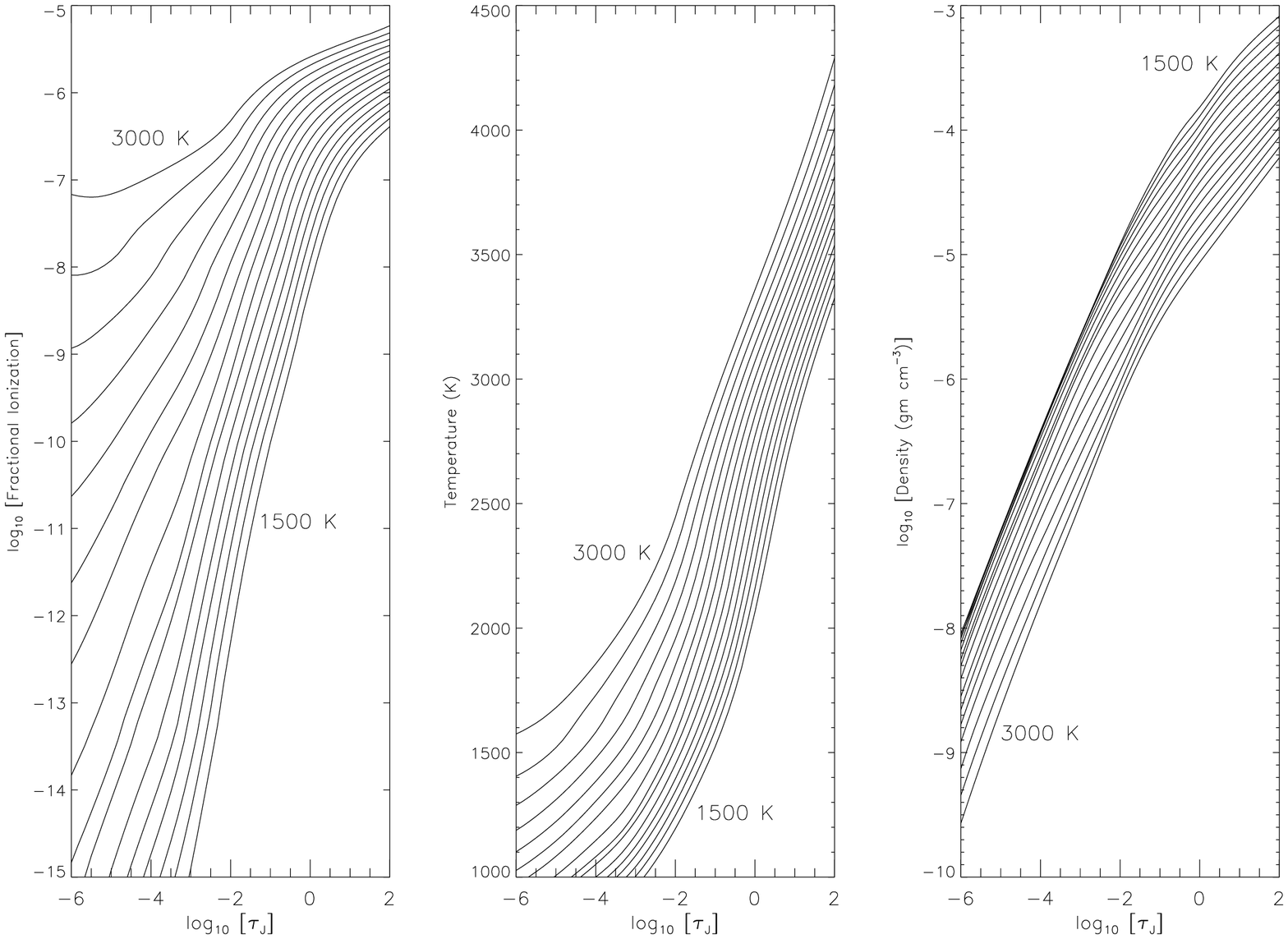}
\figcaption{\label{fig1} Various atmospheric properties as a function of ${log}_{10}$[$\tj$] (optical depth in J band). ``Cleared Dust'' AH models are used, and \teff = 1500 - 3000 K shown.  {\it Left panel}:  ${log}_{10}$[Fractional Ionization].  Fractional ionization remains very low throughout the atmosphere.  {\it Middle panel}:  Temperature.  {\it Right panel}:  ${log}_{10}$[Total Density].  The very low fractional ionizations imply that our assumption of neutral density $\sim$ total density is valid everywhere in the atmosphere.  }

\plotone{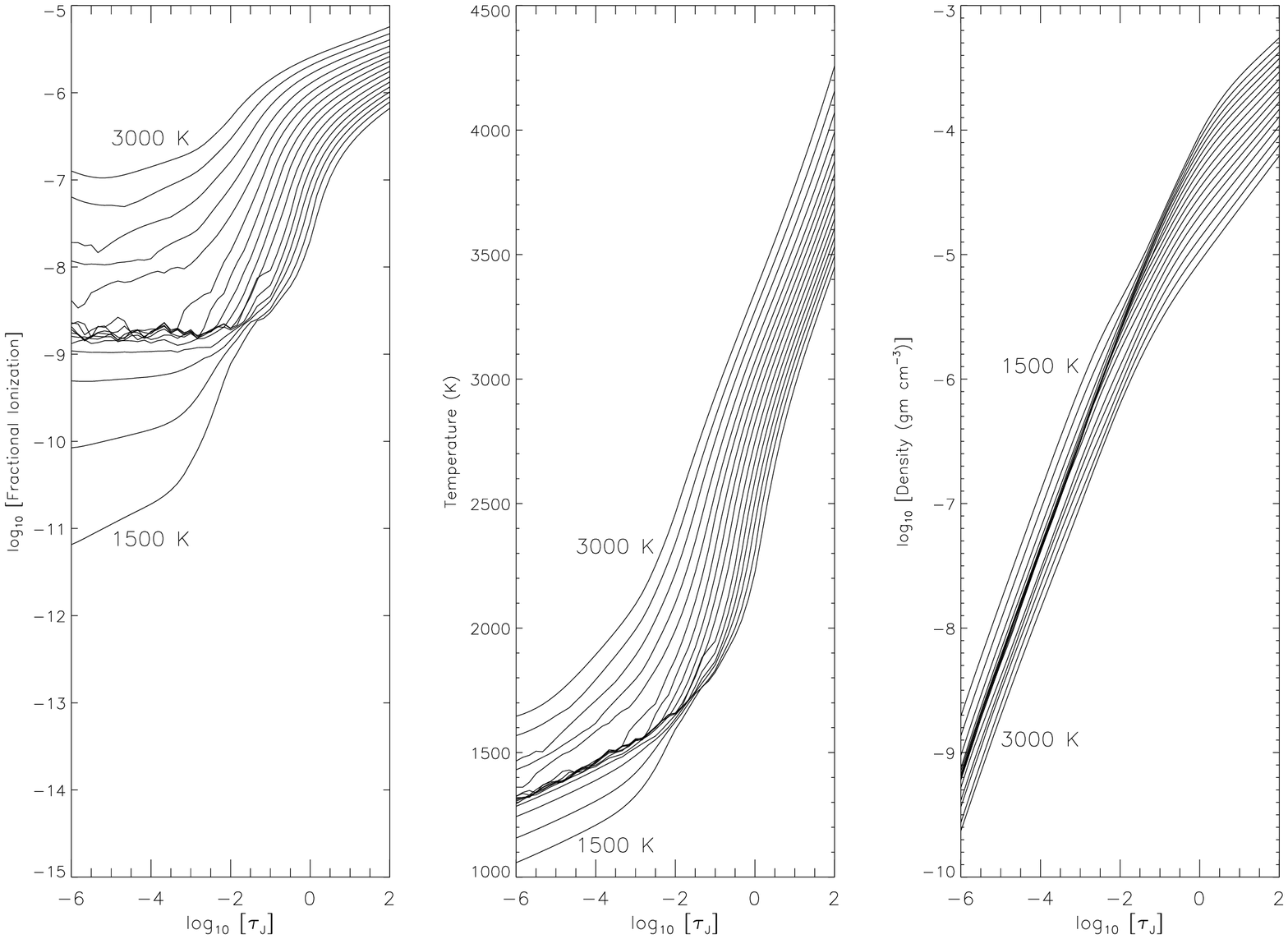}
\figcaption{\label{fig2} Same as Fig.1, but with ``Dusty'' AH models.  The fractional ionizations in the upper atmosphere are much higher here than in the ``Cleared Dust'' models, due to backwarming by dust.  However, neutral density $\sim$ total density still remains a good approximation.  In the deeper atmosphere, backwarming effects are negligible and conditions are nearly identical to those in the ``Cleared Dust'' models.  }

\plotone{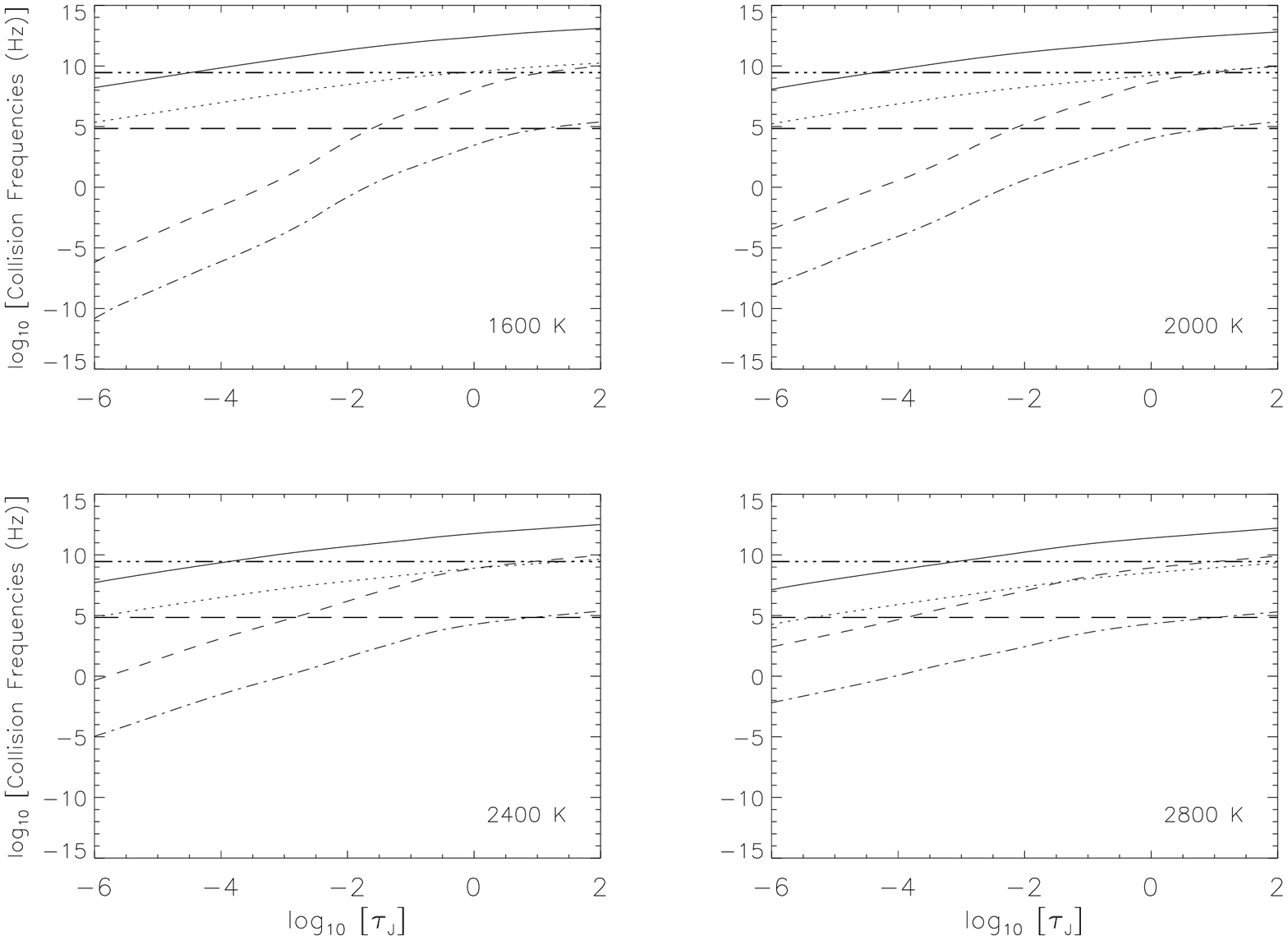}
\figcaption{\label{fig3} Various collision frequencies as a function of ${log}_{10}$[$\tj$].  ``Cleared Dust'' models are used, for \teff = 1600 K ({\it top left}), 2000 K ({\it top right}), 2400 K ({\it bottom left}) and 2800 K ({\it bottom right}).  ${log}_{10}$[$\fen$] ({\it solid line}); ${log}_{10}$[$\fin$] ({\it dotted line}); ${log}_{10}$[$\fei$] ({\it dashed line}); ${log}_{10}$[$\fie$] ({\it dash-dot}); ${log}_{10}$[$\feb$] ({\it thick dash-dot-dot}); ${log}_{10}$[$\fib$] ({\it thick long dash}).  $\fab$ is the (momentum-exchange) collision frequency for a particle of species $\alpha$ in a sea of species $\beta$, for $\alpha$ or $\beta$ equal to $i$ (ions), $e$ (electrons) or $n$ (neutrals).  $\fib$ and $\feb$ are the ion and electron gyrofrequencies respectively, in a 1 kG magnetic field. }

\plotone{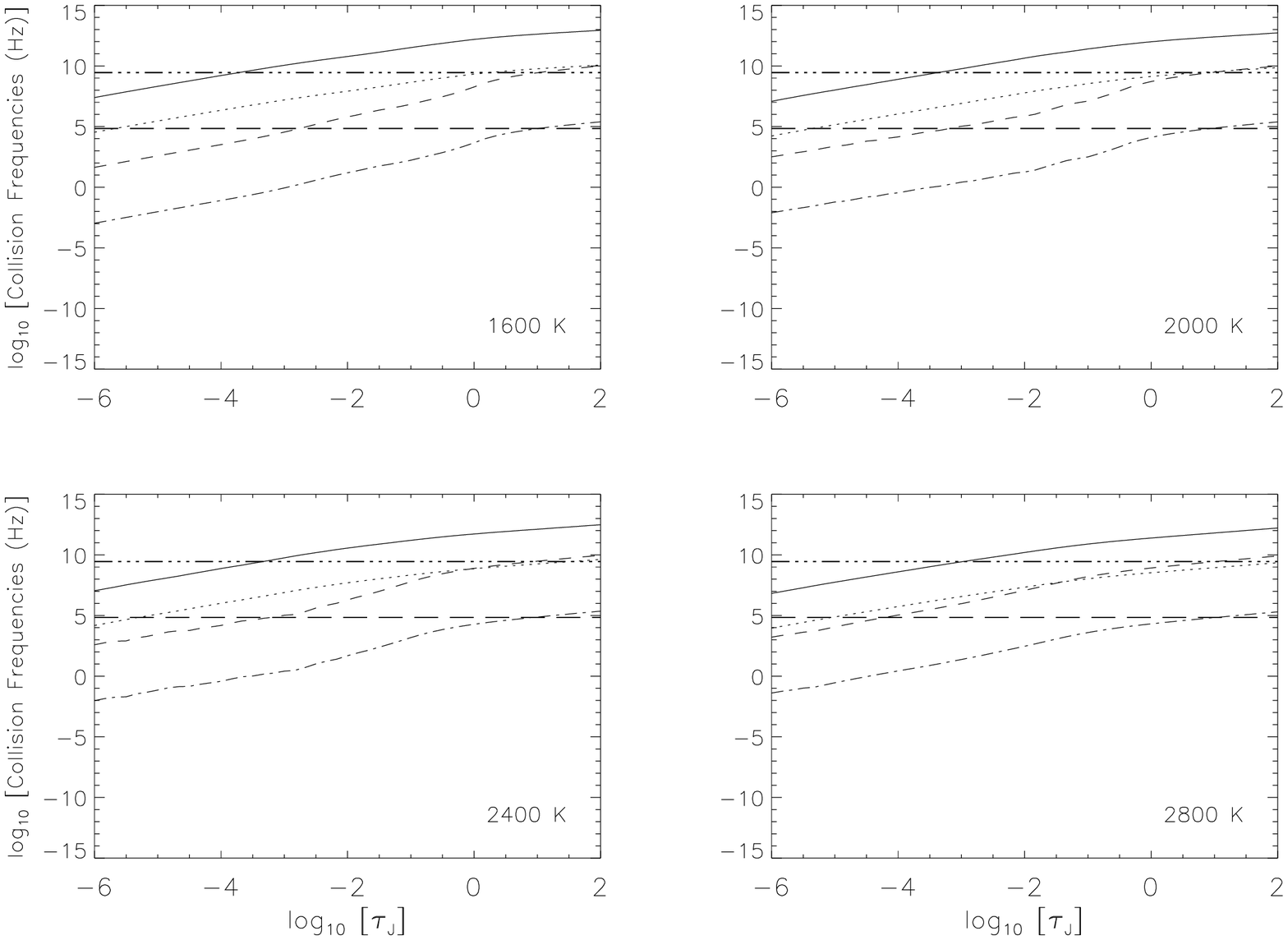}
\figcaption{\label{fig4} Same as Fig.3, but with ``Dusty'' AH models.  }

\plotone{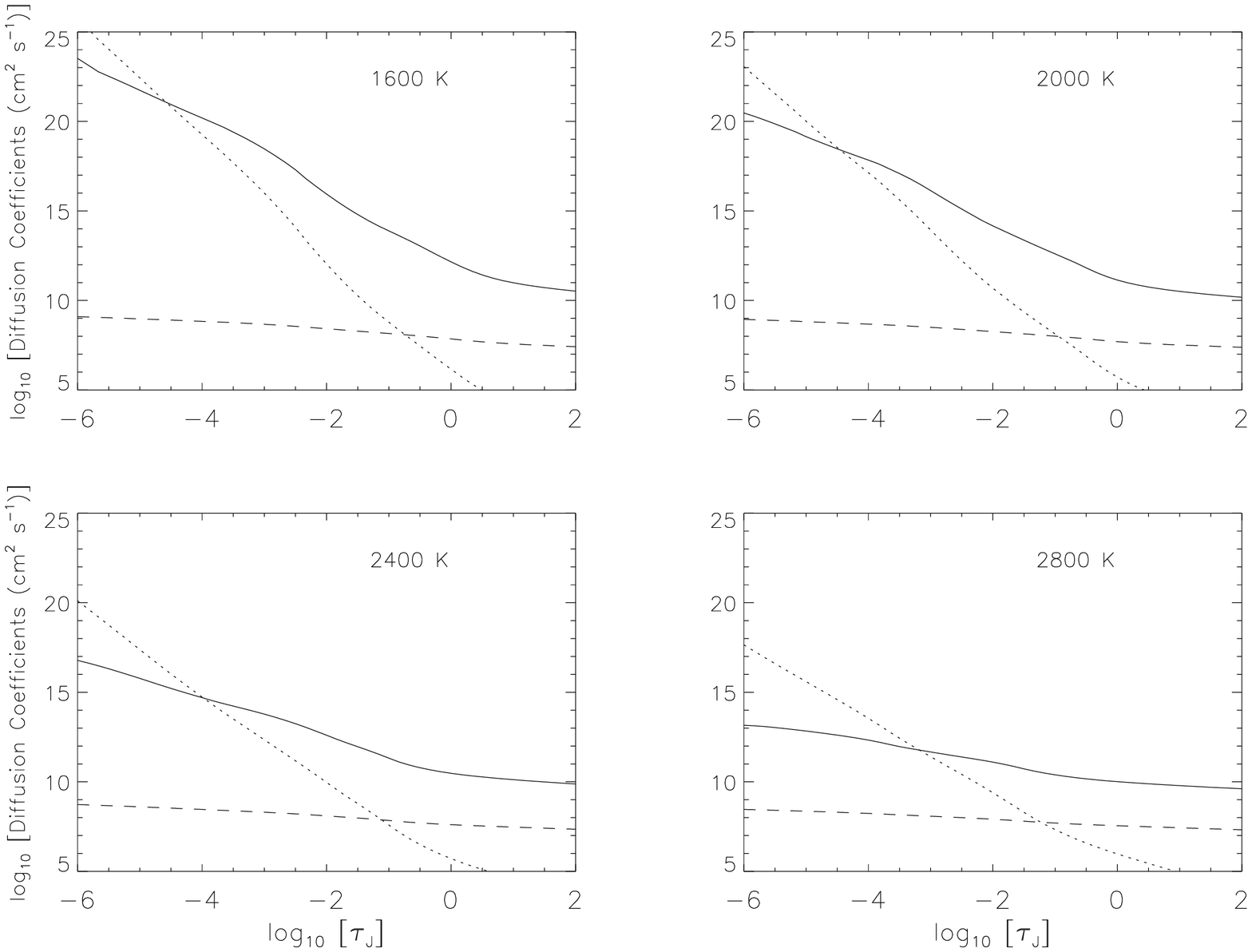}
\figcaption{\label{fig5} Diffusion coefficients as a function of ${log}_{10}$[$\tj$].  ``Cleared Dust'' models are used, for \teff = 1600 K ({\it top left}), 2000 K ({\it top right}), 2400 K ({\it bottom left}) and 2800 K ({\it bottom right}).  ${log}_{10}$[$\dcd$] (decoupled diffusion coefficient) ({\it solid line}); ${log}_{10}$[$\dcad$] (ambipolar diffusion coefficient) ({\it dotted line}); ${log}_{10}$[$\dco$] (Ohmic diffusion coefficient) ({\it dashed line}).  In these low fractional ionization (ie, predominantly neutral) regimes, $\dco$, arising from ion-electron collisions, is always negligible compared to $\dcd$, which depends on collisions of charged particles with neutrals.  In the deeper atmospheric layers, where neutral densities are high, charged particles are easily knocked off field lines by collisions with neutrals and $\dcd$ dominates, while in the more rarefied upper layers, the charged particles are better tied to the field lines and $\dcad$ dominates.  }

\plotone{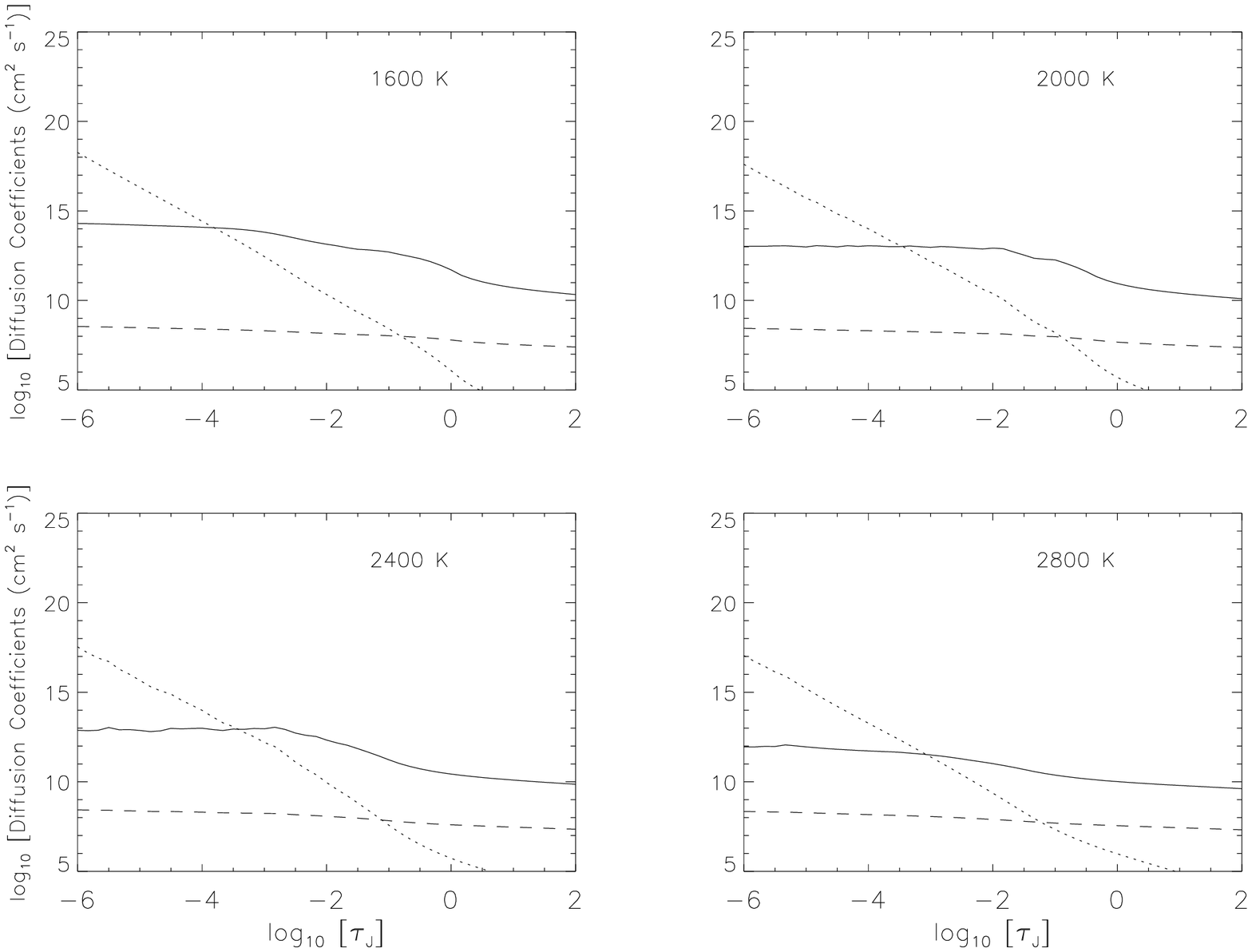}
\figcaption{\label{fig6} Same as Fig.5, but with ``Dusty'' AH models.  $\dco$ is again negligible compared to $\dcd$.  Since both $\dcad$ and $\dcd$ decrease with increasing fractional ionization, they are smaller in the upper atmosphere in these models compared to the ``Cleared Dust'' case.  Also, since the fractional ionization falls off much less rapidly with increasing height compared to the ``Cleared Dust'' models while neutral densities in both models are comparable, $\dcad$ becomes dominant deeper in the atmosphere in these ``Dusty'' models.  }

\plotone{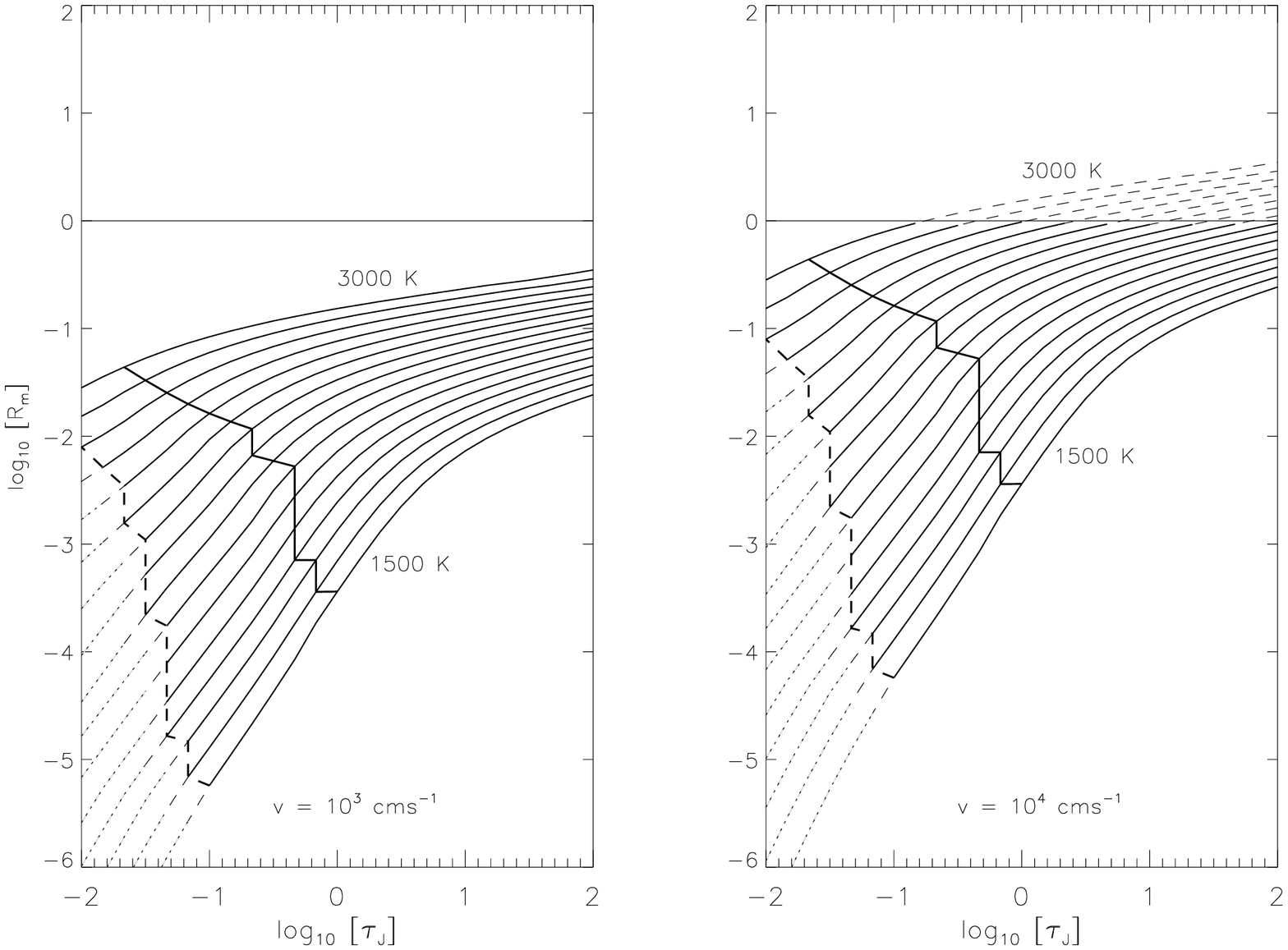}
\figcaption{\label{fig7} ${log}_{10}$[Reynolds number] as a function of ${log}_{10}$[$\tj$].  ``Cleared Dust'' models are used, for \teff = 1500 - 3000 K.  {\it Left panel}:  Using $\veff$ = 10$^{3}$ \cms.  {\it Right panel}:  Using $\veff$ = 10$^{4}$ \cms.  The {\it solid horizontal line} denotes $\Rm$ = 1, i.e., where $B_T$ becomes comparable to the background field $B_0$.  Regimes where $\B_T$ exceeds $B_0$ are shown by {\it dashed lines}; these derived $\Rm$ are not very accurate since the back-reaction of the field on the neutrals becomes important.  $\Rm$ can be assumed to be substantial in this regime, however.  The upper limit of convection according to the models is shown by the {\it thick solid curve}.  The {\it thick dashed curve} denotes one pressure scale-height above the solid curve; it indicates the lowest optical depth up to which convection may be expected, given convective overshoot and / or uncertainties in the model mixing lengths. {\it Dotted lines} indicate the $\Rm$ in regions where no convective motions are expected, if such motions did exist there.  }

\plotone{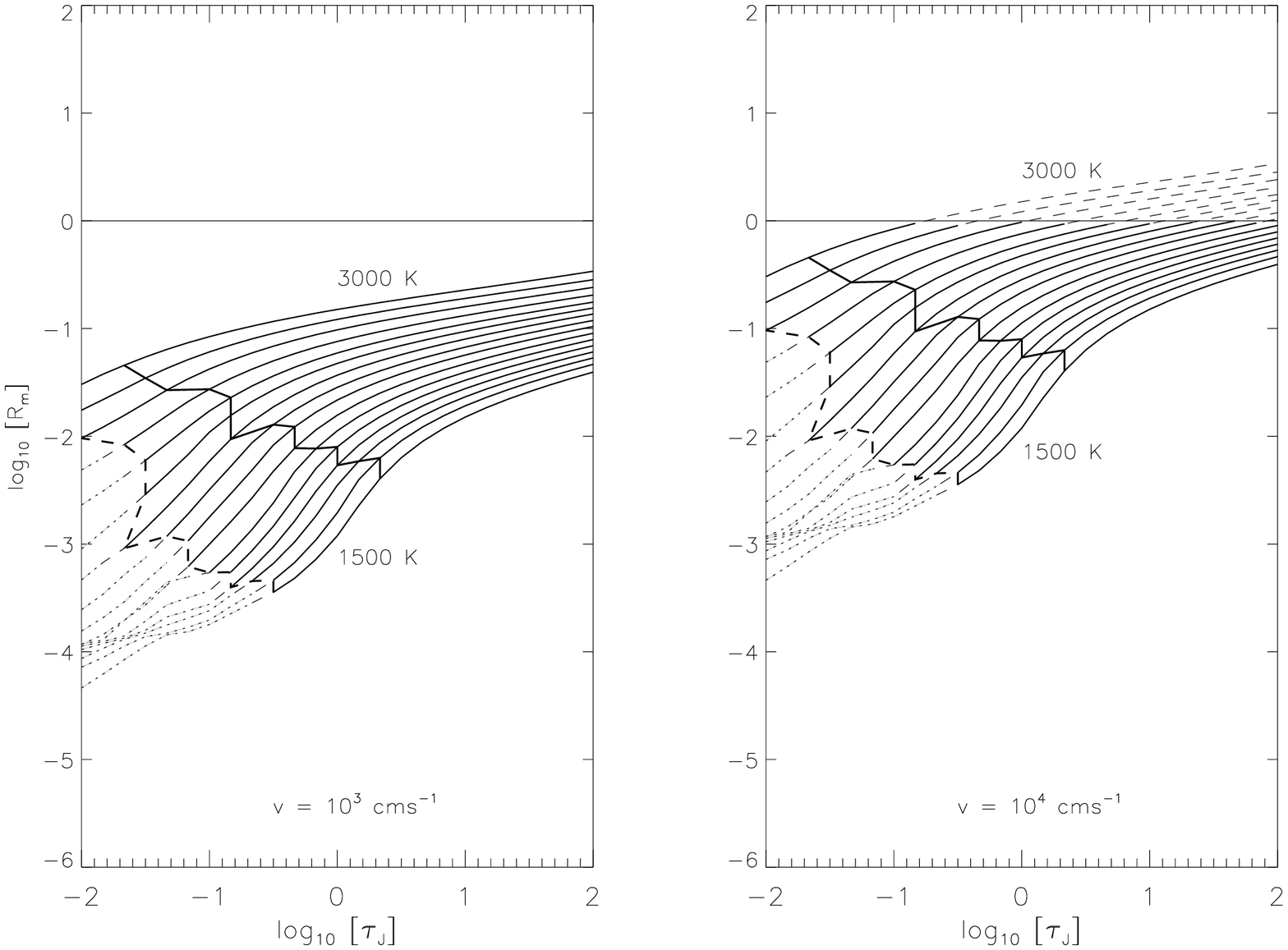}
\figcaption{\label{fig8} Same as Fig.7, but with ``Dusty'' AH models.  At low optical depths, the $\Rm$ here are much larger than in the ``Cleared Dust'' models; this reflects the higher ionization fractions, and hence smaller resistivities, in the ``Dusty'' models due to backwarming effects.  Backwarming is not important at high optical depths; in this regime, the $\Rm$ in the two models is comparable.  }

\plotone{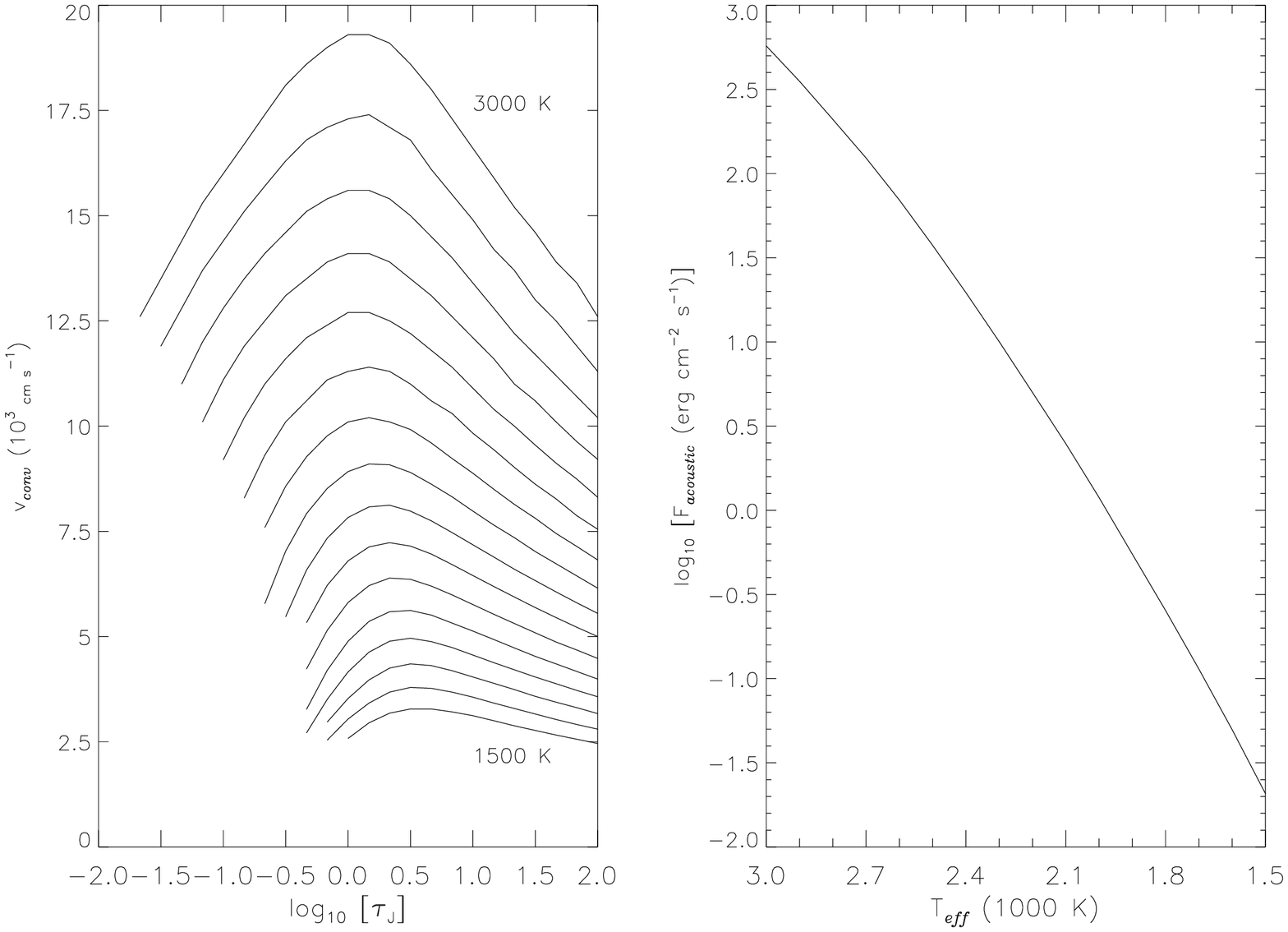}
\figcaption{\label{fig9} {\it Left panel:} Convective velocity as a function of optical depth, in the ``Cleared Dust'' AH models.  Note the decrease in velocity, at any given depth, with decreasing \teff.  {\it Right right:} Lighthill-Proudman acoustic flux versus \teff.  The fluxes are much too small to explain the observed \hal fluxes, though they do decline with \teff (like the observed \hal) as a result of decreasing v$_{conv}$ with lower \teff.  }

\plotone{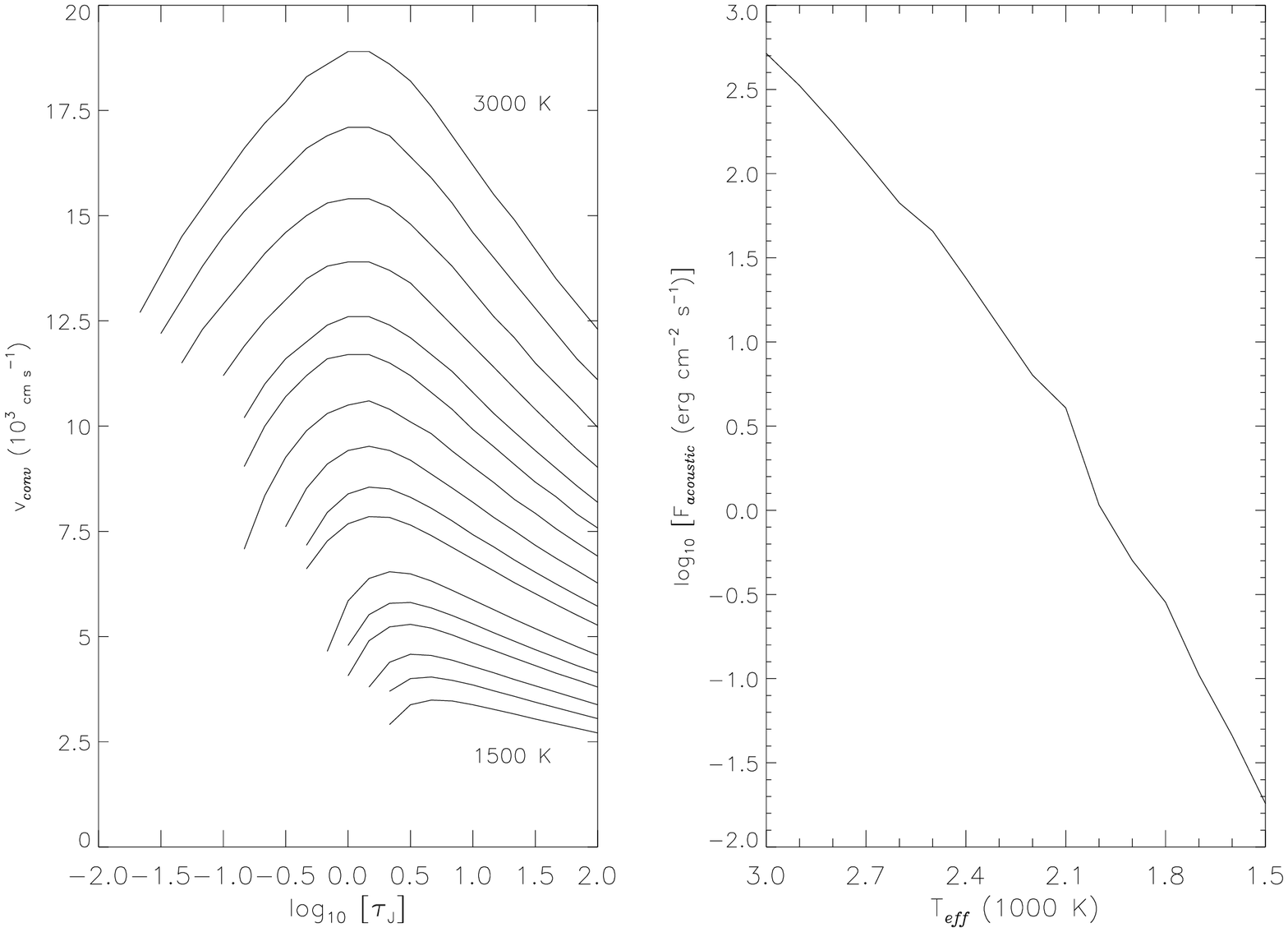}
\figcaption{\label{fig10} Same as Fig.9, but with ``Dusty'' AH models  }

\clearpage

\end{document}